# AAS Graduate Admissions Task Force: Final report summary

The goal of our task force was to produce recommendations for the AAS aimed at improving the state of graduate admissions in astronomy. We surveyed recent applicants to astronomy graduate school (Section 2) and admissions leaders in degree-granting graduate programs (Section 3), held in-depth conversations with select programs, and explored how other fields approach admissions (Section 4).

From this work we have quantified our field's evolving admissions landscape, addressed mismatches between perception and reality, and identified several key takeaways illustrating the biggest current challenges in astronomy graduate admissions:

- Students find the admissions process to be overwhelming, expensive, and stressful, with many issues caused by **poor transparency and communication** (Figure 4, 5). Since 2018 they have submitted an average of **11-12** applications per person (Figure 2).
- Graduate program applicant pools in astronomy have grown by an average of **62%** since 2018 (Figure 6). With no corresponding growth in available spots, programs have an increasingly **limited capacity** to evaluate and admit students. It is difficult for graduate programs to improve and adapt their admissions processes in the midst of a **rapidly-shifting funding and policy landscape**.
- Common themes across the astronomy community include a need for **current and accessible information** on how graduate admissions is changing and evolving across the field, interest in some degree of **standardization** or **centralization** to decrease workloads, and a desire for **guidance on best practices** to help improve all sides of the admissions process.

We have compiled four recommended actions (Section 6) the AAS should take:

| | |
|---|---|
| **Big Easy Changes**<br>Establish a AAS Committee on the Status of Graduate Admissions in Astronomy (6.1) | **Small Easy Changes**<br>Recommend standardization of application content and key communication dates (6.2) |
| **Big Difficult Changes**<br>Support adoption of a centralized "common application" system (6.4) | **Small Difficult Changes**<br>Host a set of informational admissions webpages (6.3) |

(easiest to do ↑ ; ← biggest impact on field)

Making lasting positive changes will require both immediate action and longer-term in-depth work. Our recommendations focus on achieving both of these goals.

# AAS Graduate Admissions Task Force: Final report and recommendations

*Delivered to the Board of the American Astronomical Society by the members of the GATF:*
Emily M. Levesque (chair), Rachel Ivie, Christopher Johns-Krull, Grace Krahm, Laura A. Lopez, Meredith A. MacGregor, Sebastian Monzon, Daniel R. Piacitelli, Seth Redfield, and Tom Rice

# 1 Introduction

In March 2024 the AAS formed our Graduate Admissions Task Force (GATF), asking us to produce recommendations for the AAS aimed at improving the field-wide state of graduate admissions.

The task was a timely one, as the past decade has marked some dramatic shifts in the astronomy graduate admissions landscape. The perception in the field is a stark one: application numbers are ballooning without any corresponding increase in the number of available program slots, and it is becoming increasingly challenging to run an admissions process that is fair, efficient, and effective.

Prospective students are stressed by the lengthy and exhausting application process, wrangling logistical demands that vary by school, the financial strain of application fees, and an increasingly high bar for admission. Unclear on what selection criteria schools are prioritizing, students worry about how best to prepare and present themselves to increase their odds of success. Seeing peers who apply to a dozen programs and get into none, or who need several cycles of applications to succeed, students have come to see astronomy graduate admissions as a disheartening, expensive, and difficult-to-navigate process that is nevertheless required to continue in their chosen field.

Similarly, admissions committees and degree-granting programs are overwhelmed by the challenge of fairly and effectively evaluating huge numbers of applications in a short period of time. Identifying which elements of an application are predictive of success is an immensely complicated undertaking, and efforts to be efficient are not always in line with admissions practices that avoid bias and account for inequities. Faculty who advise students can find themselves writing and submitting huge numbers of recommendation letters while also struggling to provide useful advice on navigating a rapidly-changing admissions landscape.

These perceptions are backed up by numbers. From 2018 to 2023, the number of students graduating with bachelors degrees in Astronomy increased by 57%. Graduate degree programs saw their applications increase by 62% on average and nearly 200% in some extreme cases, while the number of first-year astronomy graduate students increased by only 18%. Students apply to an average of 11-12 programs per year, and it's becoming increasingly common for students who can afford to do so to spend multiple cycles applying to graduate programs.

At the same time, the situation is more subtle and complex than it may seem. The average number of applications submitted per student has stayed remarkably constant between 2018-2023. While some programs have seen huge increases in applications, others have seen their applicant numbers stay flat or even drop. It is also very difficult to reliably track the number of students who apply to graduate programs, become discouraged by this process, and leave the field at this point in the pipeline.

After surveying applicants and admissions committees and gaining a clearer understanding of both perception and reality in the field, our Task Force has compiled a list of four recommendations for the AAS, with the goal of making a strong and lasting impact on improving the state of graduate admissions in astronomy. These recommendations are aimed at streamlining processes, increasing transparency, offering guidance to both programs and applicants, and establishing lasting infrastructure in the astronomy community to support and maintain efforts to improve graduate admissions.

As described below, the GATF surveyed both recent applicants to graduate schools (Section 2) and grad admissions leaders at degree-granting programs in astronomy (Section 3). We also conducted exploratory research in several key areas, including in-depth conversations with a diverse subset of graduate programs (Section 4.1) and investigating other fields with novel approaches to graduate admissions (Section 4.2). From the results of this work we assembled a high-level summary of the biggest challenges facing the field (Section 5) and used this to construct a set of four recommendations for action by the AAS, which we present in Section 6.

# 2 Applicant Survey

## 2.1 Survey Content, Circulation, and Response

We circulated a survey aimed at collecting information from recent applicants on their experiences and concerns regarding the graduate admissions process.

Our survey was initially dispersed on August 1st, 2024 via the AAS mailing list and further shared by members of the task force within departments and as well as through program coordinators of summer research and post-baccalaureate programs, program alumni lists, and the [Astro Grad Congress](). The survey stayed open through December 2024 and received a total of 228 responses from participants who had applied to graduate programs from 2018 onward (we specifically chose this timeline in order to include one year of pre-COVID-19 pandemic admissions data).

We collected information from respondents on their current academic standing, undergraduate degree subject and university attended, the year(s) when they applied to PhD programs, the number of cycles they applied during, and the application cycle outcomes (including number of offers, waitlists, and rejections). Additionally, demographic information was collected pertaining to age, race/ethnicity, primary language, first-generation undergraduate/graduate student, gender identity, and sexual orientation.

Regarding application content and preparedness, the survey collected information on undergraduate participation in astronomy or physics research, as well as the number of first-author/co-author publications the respondent had at the time of applications.

Finally, the survey also posed questions surveying which aspects of the application process were particularly burdensome for applicants. In addition to open-response questions asking which part of the process was most challenging and what aspect they most wished to change about the process, the survey also asked about the number of application fee waivers applied to and to what degree fees were limiting or burdensome. The full survey is available in Appendix A.

## 2.2 Summary of Results

### 2.2.1 Quantitative

**Demographics:** 228 respondents to our survey had applied to astronomy or astrophysics-related graduate programs in the US within the last 5 years; of these, essentially all had applied to PhD programs during that time interval, though 9% also noted that they had applied to Masters-only programs. 93% of respondents were either currently enrolled in or had recently completed a graduate (Masters or PhD) program when they took the survey, while another 4% identified themselves as post-baccalaureate students.

99% of our respondents had attended a US public or private university for at least some of their undergraduate studies. 70% of respondents had attended an R1 university for at least some of their education in the US, and 24.6% had attended a liberal arts colleges; however, respondents also attended R2's, R3's, minority-serving institutions, historically women's colleges, community college, Masters college, and/or a service academy. 65% of respondents earned bachelors degrees in physics, and 56% earned degrees in astronomy.

Out of 215 respondents who self-identified their race/ethnicity, 73% were white,16% Asian, 9% Hispanic/Latino, 5% Black, 3% multiracial, and 2% Native/Pacific Islander. Among our respondents 49% identified as women, 42% as men, 13% as non-binary, and 4% as transgender, while 46% self-identified as LGBTQIA+. 82% reported English as their primary language growing up, 44% were the first in their families to attempt a graduate-level degree, and 15% were first-generation college students. 90% of respondents were between 22-29 years old.

It is clear that, even with efforts to disseminate the survey broadly within the community, the respondents we reached overwhelmingly represent successful applicants to graduate school and applicants with prior educations at R1 institutions. While this offers an undeniably skewed perspective of the graduate admissions process, the responses of this group, including feedback on the application process and its challenges, are still extremely informative and worth considering closely as part of the GATF's work and recommendations.

**Application preparation:** 96% of respondents did some amount of research in astronomy or physics during their undergraduate degree: of these, 93% began research before their senior year, and 46% worked on three or more projects as undergrads. 40% of these students participated in at least one REU program.

33.5% of our respondents participated in a mentoring program or workshop designed to help students to apply to graduate school (of those, 29% were affiliated with an NSF Research Experience for Undergraduates). These mentoring programs included FAQ sessions (85%), essay workshopping (56%), advice on securing letters of recommendation (45%), and interview prep (19%).

27% of our respondents who did undergraduate research had one or more first-author publications by the time they applied to graduate school, while 44% reported having one or more co-authored publications. 83% had presented their research (as a poster or talk) at venues such as the AAS, undergraduate symposiums, or other conferences. It is important to note that students who start doing undergraduate research in their first year tend to be more slightly successful during admissions and that students who start research before their fourth year receive more offers (see Figure 1).

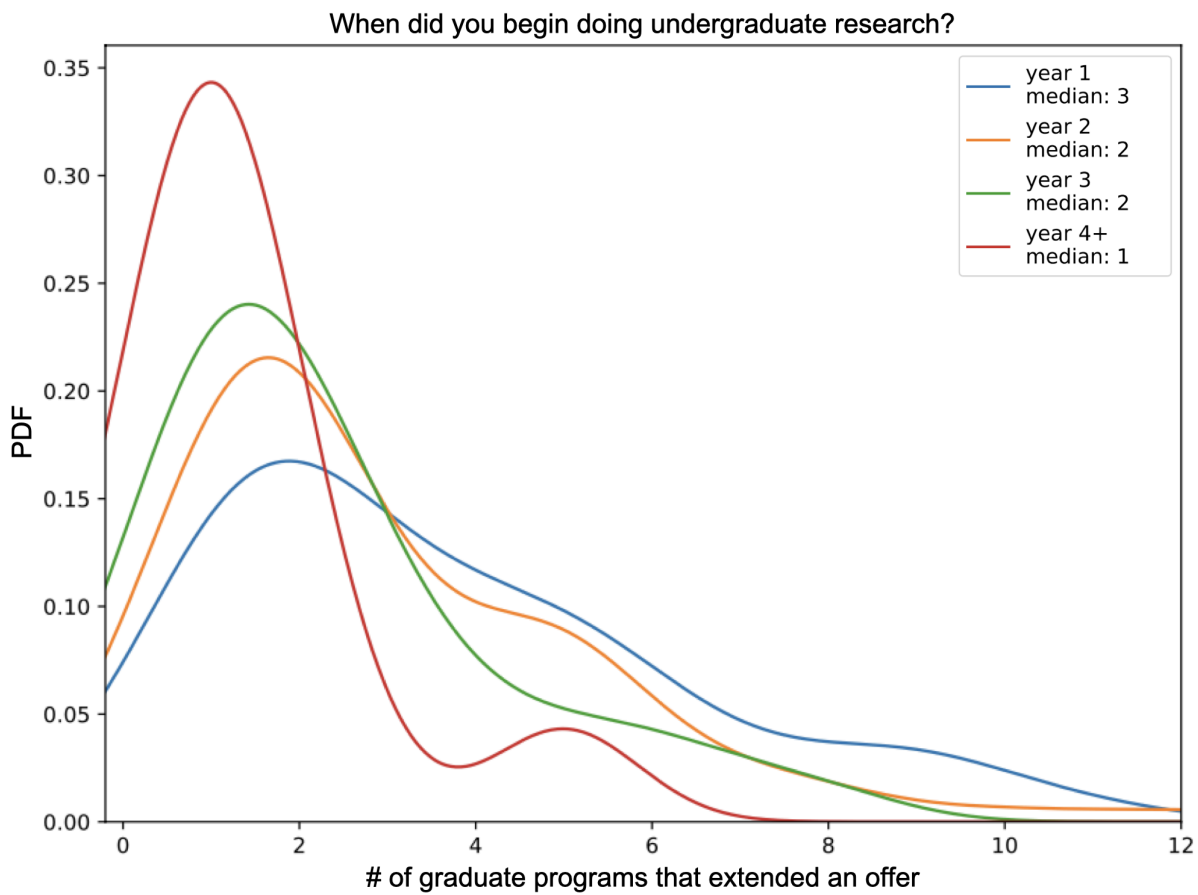

Figure 1: Probability distribution function (PDF) of how many graduate program acceptances the applicants had. These are separated into which year the applicants started research.

**The application process:**

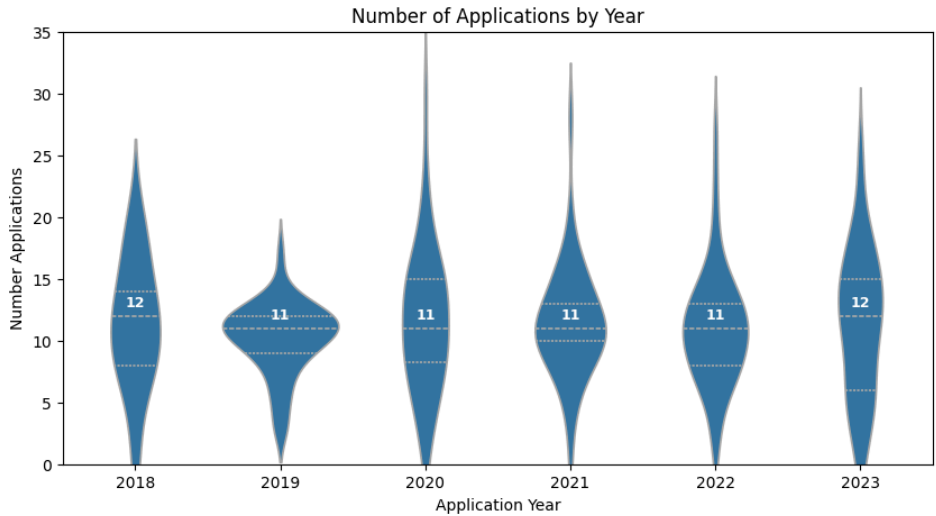

Figure 2: Violin plots illustrating the number of applications submitted per person for admissions cycles beginning in 2018-2023. The dashed line shows the median number of applications per person for each year which is labeled on each violin plot while the dotted lines show the first and third quartiles.

For the six application years that we surveyed (starting in 2018-2023), the median number of applications submitted per person stayed very consistent at 11-12 applications (Figure 2). While these varied widely from person to person (with our respondents applying to as few as 1 and as many as 35 programs), there was no increase in the median, maximum, or range of applications submitted per person over time. This implies that the sizable increase in applicants seen by many programs (see Section 3.2.1) is *not*, contrary to popular perception, caused by students applying to more and more programs. It is, however, worth noting that 86% of applicants answered "yes" (59%) or "maybe" (27%) to the question of whether they would have applied to more programs if there was a standardized "Common App"-type platform.

More than half of our respondents (53%) did not apply for any fee waivers in their most recent application cycle. However, 76% answered "yes" or "somewhat" to the question of whether the application fees were financially burdensome, while 63% answered "yes" or "maybe" to the question of whether application fees had limited the number of programs they applied to. Our program survey shows the median cost of a single application to be $75; for students applying to the average number of programs (11-12) in an application cycle this corresponds to a cost of $900.

94% of our respondents received an offer of admission in the most recent cycle; of these, 45% received 3 or more offers (Figure 3). 61% of applicants were also waitlisted at one or more of the programs they applied to. 27% of our applicants applied to graduate school more than once. 71% of our respondents applied to a single admissions cycle, 21% applied during two cycles and only 5% applied over three or more cycles. Of those that applied to more than one cycle, less than 1% received more offers the second time they applied.

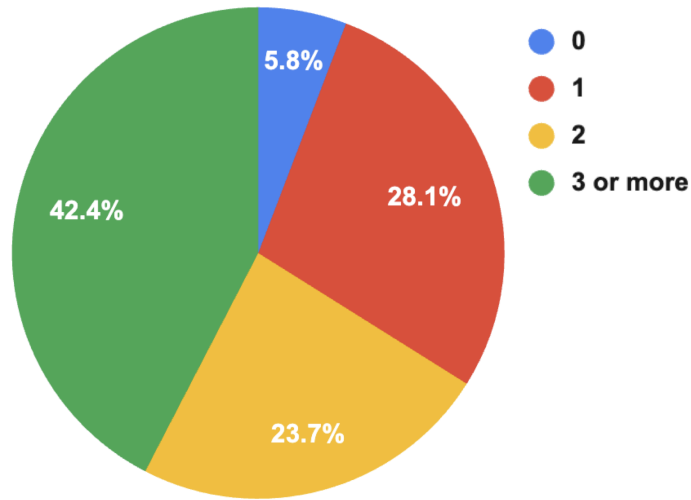

Figure 3: Number of offers respondents received in their most recent application cycle

**Application decisions:** The applicant survey also asked respondents to rate the importance of various factors in their decisions on where to apply and where to accept offers. Respondents evaluated each factor using a four-point scale: "Not Important," "Slightly Important," "Important," and "Very Important."

Across both application and acceptance decisions, the highest-ranked factors were "Faculty Research" (~93% rated it as "Important" or "Very Important"), followed by "Departmental Culture and Values" (~79%), "Physical Location" (~75%), "Long-Term Funding Stability" (~74%), and "Initial Funding Availability" (~69%). These preferences remained consistent when respondents were asked about the factors that influenced their final decision to accept an offer. These results suggest that, in addition to research availability and location, departmental climate and funding stability play a large role in applicants' decision-making processes.

Notably, ~56% of respondents rated program prestige (often perceived as a significant factor by both students and programs) as "Very Important" or "Important" in deciding where to apply and attend. While it was rare to identify program prestige as "not important" (only 7% described it as "Not Important" in choosing where to apply, while 17% described it as "Not Important" when deciding where to attend), this suggests that it is not the primary driving decision factor for most prospective graduate students.

In contrast, the factors most frequently rated as "Not Important" were "Union/Unionization Efforts" (~40%), "Telescope Access" (~40%), and "Typical Time to Degree" (~30%). The relatively low importance placed on "Telescope Access" likely reflects a mix of respondents interested in theoretical astrophysics or space-based research, for whom local telescope access is less relevant. Meanwhile, the lower importance assigned to "Union/Unionization Efforts" and "Typical Time to Degree" likely indicate that these considerations, while potentially impactful, are not primary drivers in applicants' choices.

40% of our respondents noted that they applied for at least one graduate fellowship as undergraduates; of those, 24% were successful. 52% of the applicants noted that they waited until fellowships were announced before accepting a grad school offer; while we did not ask unsuccessful fellowship applicants if they waited for fellowship news before making their decision, it is possible that a similar fraction also delayed their decisions until after large fellowships were announced (the NSF Graduate Research Fellowship Program, for example, typically announces results in early April). Notably, while 62% of recipients said that their fellowship did not ultimately impact where they chose to attend, the other 38% indicated that it either did have an effect or gave them more freedom in their decision.

### 2.2.2 Qualitative

Respondents frequently expressed a strong desire for more clear and open communications from graduate schools on topics such as admissions timelines and results, expectations from applicants, and which professors are available to mentor incoming students. Many also expressed difficulties with finding important forms and information on deadlines, particularly with respect to fee waivers and ensuring that their letters of recommendation were submitted properly and on time.

There seems to be a growing perception among applicants that publishing a first-author paper is becoming a presumed requirement, which especially impacts students who started research later in their undergraduate careers (see Figure 1). There were several suggestions of keeping or adding more non-research or academic based elements into the application process such as the GRE/PGRE, and additional ways to provide non-published demonstrations of skills, such as portfolios or writing samples (though see further discussion in Section 4.1). However, many of these suggestions were somewhat contradictory; there is no single consensus sentiment on how applicants wish to address the issue of more and more experience and training being required for admission to a PhD program.

Students from smaller and less research-focused schools in particular have noted a lack of support and instruction on their applications and have expressed interest in mentorship programs for students applying to astronomy graduate programs, similar to the [MIT PhysGAAP](#) program. Respondents also expressed interest in workshops and shared resources on applications, as well as in learning more about career paths in astronomy outside of the traditional academic route.

When asked about which aspects of the application process were the most challenging, respondents were given an open response question to detail their answers. The GATF collected, read through, and binned these responses according to commonly-mentioned topics (see Figure 4). In order of number of responses, the most frequently mentioned topics were: "Essays," "Volume of workload," "Lack of process transparency," "Personalizing each essay," and "Fees/Financials." Many respondents noted that the essay component of the application process was especially burdensome and time-consuming. While the task force recognizes that written personal statements are a key part of programs assessing applicants, we recommend standardizing the content and format of these written statements (for more see Section 6.2).

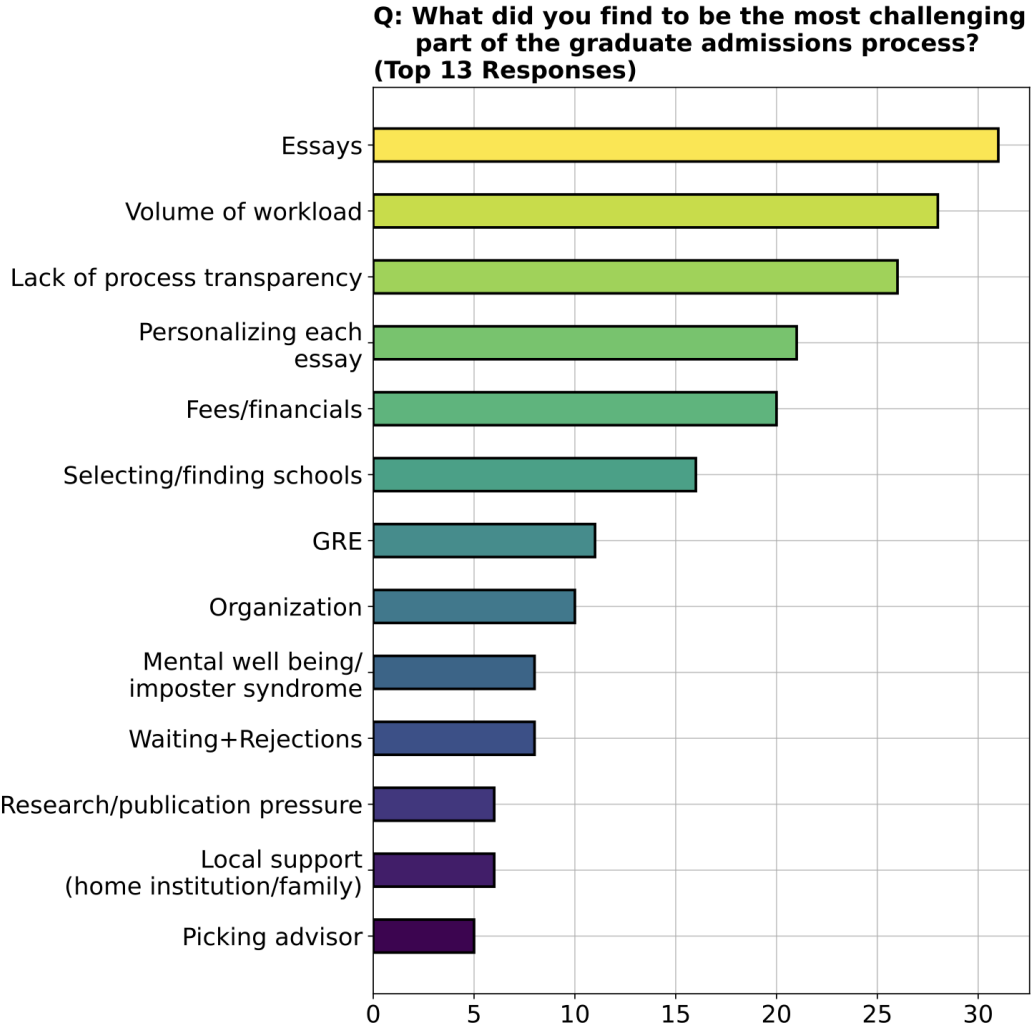

Figure 4: Binned responses for "What did you find to be the most challenging part of the graduate admissions process?"

Similarly, many responses cited "volume of workload" and the task of personalizing each essay to each program, in particular referring to the changing format guidelines from program to program (2 pages, 3 pages, 1000 words, standardized fonts and/or margins, etc.). In addition to text formatting, schools differ in what they require for statement content, such as requesting one "personal statement" and one "research statement" versus one "professional statement", often with minimal information on what these entail. With these vague and varied formats, applicants can be left unsure of what information to put in these statements (a concern also captured in "Lack of process transparency").

Finally, respondents were asked what they most wished to see changed about the application process (Figure 5). In order of number of responses, the most commonly requested changes were "Remove GRE", "Increase transparency/communication", "Reduce cost of applying", and "Adopt CommonApp." Interestingly, these responses do not precisely mirror the responses discussing the most challenging aspects of the application process: removing the GRE was the most commonly-suggested change to the process while it was only the 7th-most-frequent answer for the most challenging component. However, the majority of these responses do agree well the previous question. Most applicants hope for more communication from departments, most often in the form of wanting clear communication of their waitlist/rejection status (some applicants never heard from schools), updated program websites listing current professors and those looking to take on students, and feedback on rejected applications. Similarly, adopting a CommonApp and reducing the cost of applications also align with the most commonly reported challenges during applying.

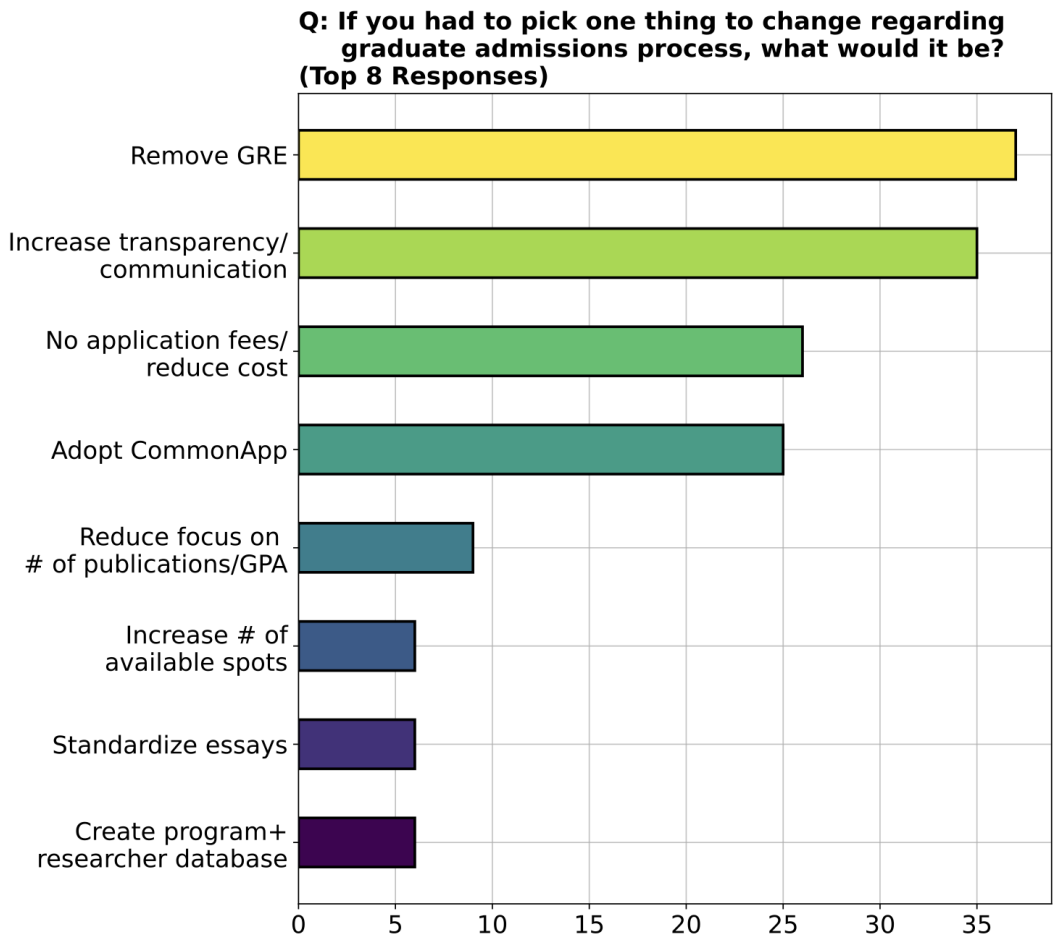

Figure 5: Binned responses for "If you had to pick one thing to change regarding graduate admissions process, what would it be?"

## 2.3 Key Takeaways

Below we summarize what we found to be the most notable takeaways based on the survey results:

**Number of Applications:** Applicants are, on average, applying to 11-12 programs every year, and this number has not increased from 2018-2023. This is notably at odds with the perception that individual students are submitting more and more applications every year, and that this is the primary factor driving increases in programs' applicant pools. While programs have indeed seen significant increases in the number of applications they receive (with an average increase of 62% between 2018-2023; see Section 3.2.1), these results suggest that the number of applications submitted by individual students is not the primary reason for this change.

**Application Fees and Their Effects:** Our surveyed applicants describe the graduate school application process as financially burdensome, with application fees limiting how many programs they applied to. However, fewer than half of the applicants applied for fee waivers. While it's not impossible that some of the impact is being overstated, this suggests that it's currently difficult for students who feel they need fee waivers to actually apply for them, and that transparency around fee waiver availability, deadlines, and criteria should be improved.

**First-author papers:** 96% of respondents to our survey did some amount of research during their undergraduate degree, and there is at least some relationship between how early students start research and how many offers they receive (see Figure X). However, it's interesting to note that, while 93% of our respondents were admitted to grad school, only 27% of our respondents had a first-author paper by the time they applied to grad school. This is at odds with the perception that a first-author paper is becoming a de facto requirement for admission.

**Standardization:** Respondents cited application essays – particularly the task of tailoring essays to match varying formatting requirements – and overall workload volume as some of the most challenging parts of the application process, along with a desire for some degree of standardization. 86% of applicants also noted that they either would or might have applied to more programs if a standardized "common-app" platform was available. While it's clear that there is interest in some degree of streamlining or standardization, and that this would improve the applicant experience, significant changes such as a full shift to a common application should be approached carefully.

**Transparency and communication:** Students expressed a strong desire for more transparency and communication during the admissions process. These included clearer information from programs (for example, up-to-date department websites listing current faculty and research interests), a better understanding of what programs are looking for in applicants, and a clear sense of when they can expect to hear from programs with updates or information on the process. Problems with transparency and communication were the most common responses across our open-answer questions asking respondents about their biggest challenges and what they most wished to see changed in the graduate admissions process.

It is worth remembering that our survey respondents represent a notably skewed view of graduate admissions. 93% of our participants were ultimately admitted to graduate school, and because of this our results lack perspectives from the many applicants who may struggle more with the process and receive no offers. However, even with this context in mind the survey results are worth considering closely: even this group of largely "successful" applicants is clearly reporting substantial concerns and difficulties with the admissions process.

# 3 Program Survey

## 3.1 Survey Content, Circulation, and Response

In addition to our applicant survey, we circulated a "program survey" intended to gather specific information on recent trends in the application process for as representative a collection of graduate programs as possible.

This survey was constructed and distributed through SurveyMonkey. It was first launched on September 26, 2024 and emailed to a list of 160 graduate programs identified as offering graduate degrees in astronomy or astrophysics; the list was assembled by combining a list of institutions maintained by the American Institute of Physics and a [crowdsourced online document](#) originally produced to track GRE and admissions fee policies in physics and astronomy graduate programs. The survey was initially emailed to department chairpersons with a request to pass the survey to individual(s) in their departments who were closely connected to graduate admissions and equipped to answer the questions, such as graduate coordinators. We ultimately received a total of 60 responses to our survey.

Our key survey goals were to get a quantitative understanding of how astronomy graduate admissions has been changing in recent years, and to distill what is most essential to graduate programs when looking at applicants so that the GATF could make specific recommendations that would streamline the process for applicants while preserving what graduate programs find most vital in making their decisions. As with the applicant survey, we requested data reaching back to 2018 in order to sample at least one full year of admissions data from before the onset of the COVID-19 pandemic.

We collected basic demographic information from departments and asked a series of questions about recent trends in the number of applications received, the number of offers made, and the number of students that matriculate each year. We also gathered information about program timelines, including application deadlines, when offers were made, and when acceptances were typically received. We asked questions related to specific policies around the application itself (including how programs communicate with applicants, whether application fees are charged, etc.), the components of the application itself (including standardized tests, letters of recommendation, student statements), how applications are reviewed, what programs typically look for in an application, and how interviews and visits are conducted. The full survey is available in Appendix B.

## 3.2 Summary of Results

### 3.2.1 Quantitative

**Demographics:** Of the 60 departments that responded to our survey, a significant majority (73%) identified themselves as R1 institutions, while an additional 20% were R2s, and 6% described themselves as liberal arts colleges. 50% of our responses were from "Physics and Astronomy" departments, 33% were Astronomy or Astrophysics departments, 20% were Physics departments, and 3% described themselves as a combination of astronomy, planetary science, and Earth science. 90% of these institutions offer a PhD.

It is worth noting that 20% of programs responding to our survey included at least some detailed answers about Masters-specific program admissions, while 7% responded with detailed answers about Bridge program admissions. Because these are relatively small numbers it is difficult to draw significant conclusions about Masters and Bridge admissions trends in recent years, but below we do note cases where distinctions between these levels of graduate admissions are worth noting.

**Application numbers:** Between the 2018-2019 and 2023-2024 application years, PhD applications increased by an average of 62%, and the average increase has been steady with time. There is, however, a significant spread in these numbers (see Figure 6). 17% of respondents who provided numbers reported *decreases* in their application numbers from 2018-2019 to 2023-2024; among departments with increasing numbers the median increase was 76%, while the largest increases reported were 193% and 195%. Application numbers for individual programs were also not necessarily linear: for example, some programs with increasing numbers of applications peaked in 2022-2023 followed by a (modest) drop in 2023-2024.

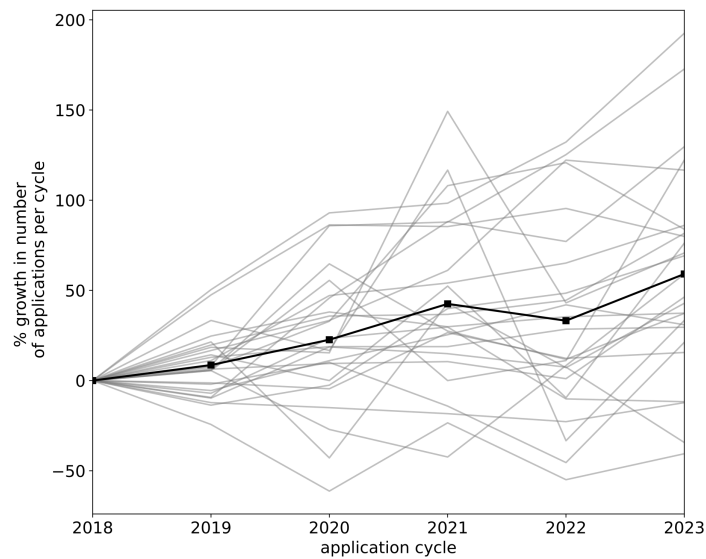

Figure 6: Percent growth in applications seen by graduate programs since 2018. Gray lines show the growth of individual programs; the black line shows the average.

By contrast, the average number of admissions offers did not change at all during this time, ranging from a low of 23 in 2019-2020 to highs of 25 in both 2018-2019 and 2022-2023. Average yields (the fraction of applicants who accepted offers) also stayed the same during this time period, with a low of 42% in 2020-2021 and highs of 46% in both 2021-2022 and 2023-2025.

Masters programs have generally seen constant application, offer, and yield numbers, ranging from 45-60 average applications per year, 10-14 offers, and 6-8 accepted offers. Bridge programs, by contrast, have grown considerably, but these programs are rare (only 4 responded to our survey) and incredibly small: our respondents described making an average of 1-4 offers per year.

**Application requirements and fees:** None of the programs that responded to our survey reported *requiring* the Physics GRE; 11% describe it as "recommended", 46% as "optional", and 43% as "not used". 80% of programs reported they had stopped using the Physics GRE between 2018-2023.

Only 5% of programs required the general GRE; another 5% described it as "recommended", 49% described it as "optional", and 41% described it as "not used". 86% of programs reported they had stopped using the GRE during or after the 2019-2020 application cycle (notably, the first cycle impacted by pandemic-related problems with taking large standardized tests).

Among programs that reported requiring recommendation letters, 82% required three letters while the other 18% required two. There was significant variation among programs that described their applicant essay requirements: departments typically requested one or two statements, with a broad variety of descriptions ("applicant statement", "statement of purpose", "personal statement", "academic statement", "research statement") and lengths typically given in pages and ranging from one to three pages in total.

88% of programs charge an application fee; these spanned from $50 to $150 with a median of $75. 77% of programs reported offering fee waivers.

**Application review:** Admissions committee sizes vary widely between programs but most (97%) draw their membership at least partially from the department faculty. While no programs reported using a "checklist" of qualifications to evaluate applicants, 44% described using a graded points-based rubric, 31% reported using an ungraded rubric or set of guidelines, and 25% reported not using any defined guidelines at all. 38% of responding programs conducted interviews as part of their application process; of these, half began using interviews in the past 5 years.

The three most important factors in grading applications (see Figure 7) appeared to be letters of recommendation and grades at undergraduate institutions (both described as "very important" or "important" by 97% of programs), as well as application essays (84% ranked these as "very important" or "important"); of these, letters of recommendation were the most highly cited as "very important", by 78% of programs. 51% of programs also ranked "reputation of undergraduate degree granting institution" as "important". It is interesting to note that, while almost 90% of responding programs noted that faculty actively communicate with potential applicants prior to the application deadline, only 14% described this type of communication as "very important", while 35% described it as not mattering at all. Finally, only 30% of responding programs rated the GRE/PGRE as "slightly important" or "important"; none list it as "very important".

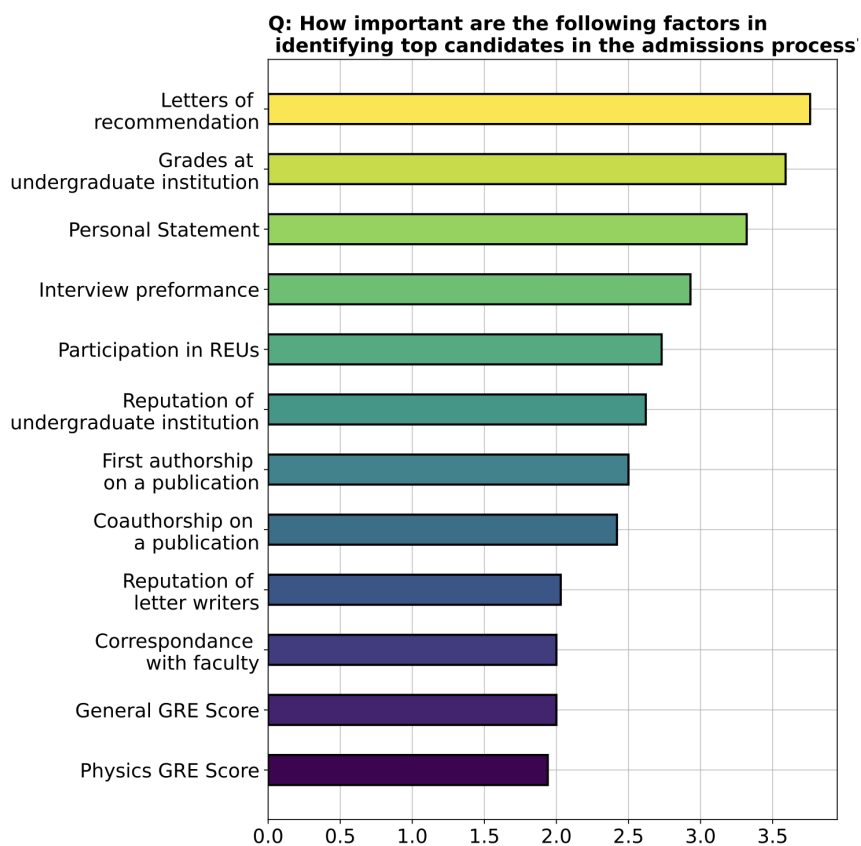

Figure 7: Binned responses for "How important are the following factors in identifying top candidates in the admissions process?"

While they were more likely to describe it as "slightly important", 89% of programs also noted the importance of participating in an REU-style intensive summer research program. Similarly, while it was more likely to be described as "slightly important" and only rarely described as "very important", 83% (86%) of programs saw first authorship (co-authorship) on a submitted refereed paper as important. However, 83% of responding programs indicated that most of their admitted students do *not* have a refereed publication.

**Deadlines and Decisions:** 87% of responding PhD programs had application deadlines that fell between early December and mid-January; 81% of programs made their first offers in January or February. 90% of responding programs offer visit opportunities to admitted students, with dates ranging from early February to late March.

When asked about the date ranges in which "a significant number" of admits notify them of decisions, 69% reported hearing from applicants between April 8-13 and 53% reported April 14 or 15 (by contrast, only 16% reported that a significant number of applicants make decisions before March 15). Half of our responding programs also noted that they extended offers after their "standard" acceptance deadline, and 33% described this as a routine practice.

### 3.2.2 Qualitative

Programs were asked to describe challenges that had stood out to them in the admissions process over the last five years. The vast majority of answers, including some Masters-specific programs, mentioned the increasing number of applicants, often combined with flat or decreasing spaces and budgets in departments. Programs also commonly described complex challenges around recruitment and yields. Limited funding, adviser availability, and programmatic balance concerns all contribute to departments making offers as sparingly as possible; however, departments often still struggle to recruit their top choices, and receiving declines from admitted students late in the process can significantly hurt yields. A number of programs also pointed to the increasingly stiff competition among highly-qualified applicants as a point of concern. Several answers, by contrast, noted problems with small numbers of applicants or recruiting domestic applicants, while others mentioned administrative challenges around evaluating, admitting, and supporting international students. The COVID-19 pandemic came up more than once as a complicating factor.

Our survey also asked departments to describe changes their programs had made in response to these challenges. The responses to this varied quite widely. Most programs addressed increases in applicant numbers by expanding or streamlining their application review process (e.g. by increasing the size admissions committee or lengthening their review timeline), although others described current or future attempts to limit applicant numbers (these included narrowing the subfields considered in the admissions process to target specific research positions, reinstating application fees and limiting fee waivers, and considering reinstating the GRE/PGRE). Several programs have recently introduced interviews as part of their admissions process, though this came with concerns about bias and how to interview effectively. Programs concerned about getting more applicants or increasing their yield mentioned increasing their program visibility and stepping up recruitment efforts. A common answer, however, was programs noting that while changes were clearly needed, none had been made yet as they were still debating the best approach.

Our last survey question asked programs whether there were larger-scale solutions (beyond their own departments) that they would like to see implemented to address current graduate admissions challenges. Many left this question blank or answered with a simple "no"; only 15% of respondents included answers with actionable suggestions. Of these, the most common suggestion involved some form of centralized application process, including a "common app"-type system, variants on the medical residency "matching" process, and community-wide implementation of things like "step-down" early decision dates (though it's important to remember that our survey explicitly asked programs about this sort of system in an earlier question; also see further discussion in Section 4.2). Several others noted the importance of understanding the significant role that financial constraints (both within individual departments and broadly across the field) play in driving graduate admissions oversubscription rates.

## 3.3 Key Takeaways

**Number of applications:** while not a universal trend, the majority of programs have seen substantial increases in the number of applications they received from 2018 to 2023. It is interesting to place this answer in context with the results from our applicant survey (Section 2.3): while programs are seeing more and more applications, applicants are, on average, applying to the same number of schools every year.

**Refereed papers:** while programs broadly note the importance of research experience, authorship of a submitted refereed paper is *not* a "requirement" for admission; 83% of responding programs noted that most of their admitted students do *not* have either first-authored or co-authored refereed publications. It is once again interesting to place this answer in context with our applicant survey results (Section 2.3), where 93% of our respondents were admitted to graduate school despite only 27% having a first-author paper. Both these results are at odds with the growing perception that publishing a first-author paper is "required" in order to get into graduate school.

**Recommendation letters:** programs overwhelmingly (97%) described recommendation letters as being an important part of their application review process, and typically require three letters. It is important to note here that the complex question of whether committees "should" use letters is separate from the practical question of whether they *do*. Letters and their use in admissions are discussed further in Section 4.1, and in our recommendations to the AAS (Sections 6.2 and 6.3 in particular).

**Challenges:** A unifying theme across the majority of our open answer responses was a sense of frustration or displeasure with the current process. However, the sources of this frustration vary from program to program and there was no strong consensus on how the situation can be improved, either within individual departments or throughout astronomy as a whole. The most common suggestions pointed to some degree of standardization or centralization, either explicitly mentioned or through requests for introducing field-wide policies.

# 4 Exploratory Research

## 4.1 The "Deep Dive" conversations

While our program survey gave us a valuable overview of the current graduate admissions landscape in astronomy from programs' perspectives, it was still an incomplete measure of the broad range of admissions goals and challenges faced by graduate departments. Beginning this research, we were keenly aware that astronomy graduate admissions can look extremely different based on department structure (astronomy can, for example, be a stand-alone department, paired with physics or planetary science as a distinct entity within a shared department, or a subdivision of a larger physics or space science program), terminal degree (MS or PhD), host institution research focus (e.g. R1, R2, SLAC), size, geographic area, and sundry other factors that can alter how a program approaches its admissions process.

With this in mind, we supplemented our program survey with a series of what came to be known as "deep dive" conversations, where members of the GATF sat down with admissions leaders in nine different departments. Combined with the programs represented among GATF members, the final list of programs that we were able to discuss in-depth spanned a mix of public, private, R1, R2, and SLAC institutions, Masters- and PhD-granting programs, departments (including Astronomy, Physics, Physics & Astronomy, and partnerships with planetary/space science fields), and all five major geographical regions of the US (Northeast, Southeast, Midwest, Southwest, and West). This allowed us to explore the details of the admissions processes and challenges at a broad range of institutions. We spoke with all programs on the promise of confidentiality: our goal in this report is to share what we learned from these conversations in a general and informational sense rather than quoting institution-specific concerns or referencing clearly identifying details.

In each conversation we asked department representatives three primary questions. First, we were interested in what they were happy with, or felt was working well, in their admissions process. Second, we asked them to list and elaborate on some of the key challenges they were currently facing with admissions. Finally, we asked how they thought the AAS could be helpful to them in the admissions process, including preliminary discussion of some recommendations the GATF was considering. We summarize some of the most common and striking responses to our questions below.

**Application strength:** The programs we spoke with were unanimous in commenting positively on the high caliber of applications they received. Every program stated that the ultimate outcome of their process from year to year was excellent, in spite of logistical challenges, and attributed this at least in part to the notable depth of talent in their applicant pool.

Many programs did, however, state concerns about fairly and effectively evaluating such a strong set of applications. Several specifically cited concerns with the idea of credential inflation: a common refrain from interviewees in our conversations was some variation on "I'm glad I'm not applying to grad school today", noting that the bar for standing out among peers has become quite high.

One competitive program within a larger physics department noted that a significant fraction of their admitted class in the previous year had a first-author paper on their CVs, and mentioned this alongside the sentiment that there are "many different ways to be successful in astronomy" and a concern that an applicant "arms race" would make it difficult for strong but unconventional applicants to stand out, or to rise to the top as part of their larger physics process.

Another program raised the concern that, particularly when it came to grade inflation and increasingly common undergraduate research experience, it was becoming hard to differentiate between applicants and rigorously assess their academic preparation. Both of these sentiments echoed comments received in our program survey open answers (see Section 3.2.2): one answer noted that inflated grades and uniformly-positive letters of recommendation made it difficult to form a realistic picture of applicants, while another raised concerns that the field had become so crowded that only "perfect" applicants were now being admitted.

Finally, while our interviews were mainly focused on an admissions committee perspective, most participants also volunteered their experiences as instructors or advisers trying to guide students through the process. These comments often included frustration that acceptance rates seemed to be dropping for even their strongest applicants and concerns that the bar for what should be considered a "stand-out" application was becoming extremely high.

**Number of applications:** Every program we spoke with had seen a substantial increase in their applicant pool in recent years and commented on this as a significant change that had impacted their admissions process: the smallest program we spoke to received ~50 astronomy-specific applications within a physics & astronomy department while the four largest received upwards of 400, including both astronomy-specific programs and astronomy-focused applications within a larger physics department. Several programs that had seen sudden sizable increases (as high as ~200%) pointed specifically to changes in their GRE/PGRE policy, noting that their applicant number spiked when they removed GRE/PGRE requirements. Notably, one program, a physics & astronomy program at a public R1 in the Midwest, was pleased to have seen a doubling of their applicants in recent years and saw this as a good sign that they were successfully attracting more and stronger students. Another, an astronomy-specific program at an R2, was similarly happy with their application numbers, noting that they felt they were attracting a good number of highly-qualified students with interest in the specific subfields focused on by their program.

Most programs were concerned about the logistical and workload challenges that came with this increase, and several noted that their current process was at or past capacity. Some noted that, even with the aid of rubrics, a holistic review that involved giving every application a careful and detailed read was becoming impossible due to the sheer volume of applications. Three different departments specifically noted their small and static number of faculty members, highlighting that this makes them unable to increase their committee size to accommodate growing applicant pools.

Even in departments that had attempted to increase their committee size, recruiting enough faculty to match the increased workload was difficult. Several programs specifically mentioned the challenge of "churn" on these committees, with individual members serving for only one year before rotating off and citing the high workload as a reason. By contrast, one astronomy-specific program at a large public R1 university noted that they involved a significant fraction of their faculty in the admissions process from year to year and felt that, as a result, this encouraged "faculty buy-in" and a department that was largely pleased with the efficacy of their admissions process.

Two programs that were notably similar in profile (private high-prestige R1 astronomy-only programs on the East Coast) had notably different experiences of their growing applicant pools' demographic impact: one noted that the increase in numbers had significantly broadened the diversity of their applicant pool, while the other expressed concern that this growth had not led to significant changes in both their applicants' and their admitted students' diversity, leading to worries that other factors might be preventing students from applying. The latter program noted that, despite an already-substantial increase in application numbers, they were eager to find ways to remove obstacles to applying and "ready to handle it" if this meant their applicant pool continued to grow.

**Evaluating applications:** Eight out of the nine programs we spoke to used a rubric as part of their evaluation process, and half of those programs explicitly cited the rubric as an element of the process they were particularly happy with. These programs noted that a detailed rubric with clear categories and scoring guidelines made evaluating applications both more streamlined and more effective. While programs' rubrics and how they used them varied, several mentioned AAS workshops or other publicly-available resources as a valuable means of finding and refining their use of rubrics. One program did raise concerns about rubrics offering a means for committee members to "abuse" the numerical system, using numerical weighting to obscure personal preferences, while several others noted that their rubrics were still being tweaked and that they expected to see some changes in current and future years.

Our programs' committees all included faculty members; however, several also involved current graduate students at various steps in the process. Grad student participation ranged from participating in the interview stage to full membership on the committee until the final voting stage. These programs largely found student perspectives valuable, and the degree of student participation was mainly limited by official university policies; however, several programs that limited student participation did cite concerns with application confidentiality or not wanting to overload their department's graduate students with service responsibilities.

Several programs expressed concern that admissions was becoming more subjective, and that removing one element of an application – typically the GRE/PGRE – put more weight on other elements that merely introduced other biases or problems, such as letters or access to research experience. More than one program had either reinstated a GRE or PGRE requirement or was actively considering such a change, and typically cited a desire for more quantitative or straightforward admissions data as the motivation for the change.

Similarly, programs that conducted interviews found them extremely valuable (noting that it helped them both to select strong applicants and to forge early connections that were helpful in the recruitment process) but raised concerns about whether the interviews were introducing problems. More than one program expressed worry that interviews could unfairly select for charismatic applicants, or that they lacked guidance on best practices for interviewing.

Transcripts were cited as an important but hard-to-use element of many applications: most programs request official transcripts from students' previous institutions, which can be expensive for students to provide and are typically delivered in a wide variety of different and difficult-to-read formats. The simple challenge of reading and parsing transcript information to evaluate, for example, applicants' performance in upper-level physics and astronomy classes, was brought up by several programs as a minor but time-consuming concern.

Finally, several programs volunteered that their evaluation process was slowed down by difficulties with their larger departments or institutions. Astronomy-focused evaluators in a physics department felt that they were at odds with colleagues in other subfields of physics when it came to using rubrics and evaluating applicants, while other departments struggled with centralized application systems required by their universities that did not effectively handle large numbers of applications or placed significant administrative burdens on department faculty and staff.

**Recommendation letters:** letters in particular were often mentioned as both an important and a problematic part of the application review process. While all the programs we spoke to required letters and saw them as a significant part of their evaluation, concerns were raised about letter "inflation" (with recommenders writing increasingly-superlative letters that make them difficult to interpret); excessively long or detailed letters; poor opinions of Likert-style ranking buttons (asking writers, for example, to rank individual traits or state whether an applicant is in the top 1%, 5%, 10%, etc. of students they have worked with) that were nevertheless often required by college or university letter submission interfaces; and worries that the fame or familiarity of a letter writer might be unfairly weighted. One program simultaneously noted that reading multiple letters was a useful way to identify students that had good research and collaborative skills, but that these letters were often excessively long and detailed and unnecessarily tailored to their program.

It's worth noting that in late 2024 an open letter, "*On the Use of Letters of Recommendation in Astronomy and Astrophysics Graduate Admissions*" (Barron et al. 2024) was posted online by a collaboration of early career faculty; the letter's abstract mentioned that the intended recipient was the AAS GATF. In it the authors enumerate several structural problems with the current way that astronomy graduate admissions uses recommendation letters. These include a substantial time and effort burden for letter writers (who often write letters for multiple students, navigate a myriad of different admissions portals and ranking scales and deadlines), the well-documented potential for these letters to propagate bias, and the sizable reading burden that letters place on admissions committees. The open letter also questioned the efficacy and practical use of letters, noting that letters are often ultimately used to spot "red flags" for individual students, contextualise the students' described research experience, or to get feedback from known and trusted colleagues that can be strongly weighted when considering an application.

As an alternative to the standard application requirement of three recommendation letters, Barron et al. (2024) proposed a hybrid model that combined a student-submitted portfolio with one or two (rather than three) letters. In this proposal students would submit research-relevant work products (ranging from posters or writing samples to code samples, videos, or graphics) and shorter "context statements" (the letter gives a 1000-character limit as an example) from one or two different advisers to a centralized platform. It's worth noting that this model does still include "letters", but that the letters are notably shorter and specifically aimed at addressing the students' research experience, removing the "teaching letter" component but adding a significant and diverse array of potential "portfolio" materials for admissions committees to evaluate.

The programs that we spoke to described substantial concerns with existing committee workloads, but did not cite letters in particular as a burdensome amount of reading and did not express a desire for more materials from applicants in the form of "portfolios" (several interviewees noted in conversation that when applicants independently provided extra applications materials they were most often ignored due to sheer time constraints). Only one of the programs that we spoke to used a version of a research portfolio as part of their application, requesting that applicants who had made it to the shortlist/interview stage submit a sample of their scientific writing. It's worth noting that this sample was requested for only a small subset of applicants, and that the program still used standard recommendation letters in the earlier stages of their process; however, they did specifically cite this scientific writing sample as a useful part of their evaluation process.

However, several programs did note that they would like to see shorter or more tightly-focused letters. This included discussion of what were seen as relatively ineffectual "teaching letters", where applicants request one of their three letters from professors who know them only through relatively impersonal coursework rather than through research or other one-on-one interactions. The Barron et al. (2024) suggestion of fewer, shorter, and more-tightly-focused research letters agrees well with this feedback.

**Timeline:** three programs with large (200+) applicant pools specifically mentioned the decision timeline as a significant problem in their admissions process. All of these programs used waitlists, but noted that many of their prospective students did not make final decisions until the final 72 hours before the April 15th deadline. In cases where their prospective students declined this often led to a chaotic and rushed process of reaching out to waitlisted applicants (some of whom had already accepted offers elsewhere) and requiring decisions on a relatively fast timescale. Other programs that did not employ waitlists did not share the timeline concern but similarly noted that decisions from students often came very late in the process, regardless of whether the decisions were acceptances or declines.

**Funding and programmatic balance:** Four different programs expressed concerns about how best to handle the challenges of programmatic balance and funding opportunities within their departments. As one program described, a common debate from year to year was the issue of whether to "only admit students who want to work with faculty [who have funding]" or to simply "admit students who are great and…figure it out". From year to year programs often need to balance available funding, concerns about programmatic balance (admitted students with interests that agree with both current and future research needs), and logistical challenges such as available student office space.

Another program dealing with similar challenges noted that the basic need to match incoming students with available funding also risked over-optimizing on students' stated field of interest in their applications: as they noted, students "shouldn't have to know what they want to do" as incoming first-year graduates, but giving students the flexibility to explore projects is becoming increasingly challenge as available research funding at universities becomes increasingly limited and uncertain.

A related concern was information asymmetry on the applicant side: programs noted that applicants often don't know which faculty members are taking research students from year to year, with challenges ranging from out-of-date department websites that list retired professors who don't maintain active research groups to unclear information on which advisers have available grant funding. This limits applicants' abilities to choose which programs to apply to and makes it difficult for programs to match incoming students with funded research positions. Some programs discussed or experimented with putting clearer information about available research areas on their department websites, but while this gave applicants more information it didn't solve the fundamental concern on the departments' end of how, or even if, they should go about matching student and faculty research interests at the offer and recruitment stage.

**How the AAS can help:** the most common response to our question of how the AAS could help improve graduate admissions, mentioned by six of the nine programs we spoke to, was a desire for some standardization of application content, with a particular emphasis on simplifying and standardizing recommendation letters. Most programs mentioned some caveats or limitations to this – noting, for example, that even in a more standardized application they would still want applicants to clearly address why they were specifically interested in a particular program – but speculated that standardized application content would improve both student and committee workload.

Five programs also mentioned that some "best practices" recommendations from the AAS on, for example, the use of rubrics or how to effectively interview would be very helpful. In addition to the straightforward content of these recommendations, three programs (that notably include an astronomy department, a physics and astronomy department, and a physics department) explicitly noted that the authority of AAS-backed recommendations would help with encouraging their colleagues to adopt any proposed changes.

Several programs also mentioned interest in some aspects of what they colloquially referred to as a field-wide "matching" system. These discussions were based on general impressions of the medical residency application process and often focused on the appeal of either a centralized application system, a standardized common app, or some more transparent means of matching students and programs based on available space and interest. We discuss these approaches in more detail below (Section 4.2).

Finally, a common refrain throughout our conversations with admissions leaders and committee members was significant concern for the challenges faced by their applicants. Our interviewees recognized that students found applying to grad school to be stressful, expensive, and time-consuming, and many expressed a desire to improve both the student experience and the overall fairness of the graduate admissions process.

## 4.2 Approaches of Other Fields

### 4.2.1 Common Applications

In response to large and growing numbers of applicants, higher education has turned to common application systems. At the largest (and thus most widely known) scale are the common application systems associated with entry into undergraduate degree programs in the United States. Organized by several non-profits and state-run university systems, this approach serves over one million applicants each cycle and is utilized by over 1000 institutions. The most widely used are the Common App and Coalition App, although several state university systems (e.g., California, Cal State, Texas, SUNY) have also implemented their own versions of the common application for prospective students, individually serving >250,000 students each cycle.

An undergraduate common application system that is closer to the scale of astronomy is the QuestBridge program. QuestBridge, structured as a 12-choice matching system, is primarily focused on low-income students applying for undergraduate programs and matches them with scholarships at specific institutions. While the program is undergraduate-focused, it does partner with graduate programs in business and medicine to offer alumni scholarship matching opportunities. While the structure and mission of QuestBridge is different, its scale and approach may provide an interesting model.

Perhaps the most frequently mentioned graduate degree that involves a common application is entry into medical school and, later, entry into a residency program. The scale is still significantly larger than astronomy, with >50,000 applicants per year, and the program has been in place for about 60 years.

Medical school applications are administered by a national association (the Association of American Medical Colleges, or AAMC). It has a common application platform but applicants also need to submit a supplementary customised application for each institution. Letters of reference are uploaded to this common platform, creating a "portfolio" of recommendations that can be utilized by all institutions (undergraduate institutions may even submit summary letters drawn from the recommendations of several faculty, and well-resourced institutions may have significant infrastructure on campus to facilitate this for their students). The costs are significant and all entities administer a fee: applicants pay $175 to submit a common application to the first school and $46 for each additional school, and individual institutions charge $50-$150 for their supplementary application. The AAMC also monetizes their data and information on the application process and sells that to applicants for $30. Students typically apply to 15-20 schools. There is no matching element at this stage, but the process does have a down-select stage: applicants that are admitted to many schools can only retain 2-3 of them about 2 weeks before the formal acceptance deadline.

### 4.2.2 "Matching"

The medical residency application process, administered by the non-profit National Resident Matching Program (NRMP), is slightly different from the admission process to medical school. It utilizes a proprietary two-set matching algorithm, or Gale-Shapley algorithm, to optimally combine ranked lists from both applicants and programs. Applicants apply to programs through another common application platform, the Electronic Residency Application Service (a service administered by the American Association of Medical Colleges). Programs review applicants and invite interviews. After the interview stage programs and applicants both compile rank-order lists that are submitted to the NRMP. The algorithm then matches applicants and programs, initially prioritizing placing applicants with the program most preferred on their list.

The infamous "Match Day" associated with residency matching was somewhat overhauled after 2010. Now, on Monday of Match Week applicants in the program learn if they have matched, but not where. A list of unmatched applicants and programs with remaining slots is then released, kicking off a series of very rapid-fire rounds over the course of the week in which programs with open spots can make offers to unmatched applicants (this is the Supplemental Offer and Acceptance Program, or SOAP, an updated version of the process previously known as "the Scramble"). Finally, Friday of Match Week is Match Day, when the full results of the algorithm are released and prospective residents learn where they have matched.

It is important to note that the residency matching process happens later in students' careers, on a much larger scale, and requires substantial and dedicated support structures at every step of the process. The first step of the process is also still very much a "classic" admissions procedure, with programs reviewing applications and conducting interviews, and the subsequent matching stage relies heavily on a uniform timeline across the field. Finally, the matching algorithm places enormous weight on applicants' ranked lists and the process allows very little flexibility for applicants who may change their minds; as a result, applicants must be very well-informed about their prospective programs and confident in their choices.

The idea of a matching process comes up frequently in discussion of graduate admissions, and applicants and programs alike are intrigued by the possibility. However, it is difficult to envision how a closely-analogous process to residency matching would be implemented for US astronomy: the cost and personnel requirements, infrastructure overhaul, and methods for how applicants and programs evaluate one another are all at odds with the current state of astronomy graduate admissions and would require an immense investment of time, effort, and funds to implement.

While not a matching program in the large-scale algorithmic sense, some natural science fields have adopted a model where applications are more directly matched to the research mentorship and funding of a specific faculty member or research group. Here, faculty who are interested in and capable of supporting a graduate student go to the pool of applicants and select their top choices. This model encourages applicants to build a relationship with potential advisers prior to submitting their application. This is significantly different than how astronomy has traditionally developed, with a more open-admit format administered by the entire faculty or a committee of the faculty (however, as discussed above, in practice many astronomy programs implement at least some version of this process, even at the simplest level of admitting students with some consideration of how their interests align with faculty members or groups that have funding available). The implementation of this approach also varies considerably from program to program and even from year to year, and variations in, for example, whether and how rubrics are used and how individual relationships are forged, can pose problems for programs that wish to prioritize fairness, transparency, and efficiency.

### 4.2.3 Takeaways for Astronomy

Many of the admissions approaches presented above have existed in some form for many decades in other fields. These fields have devoted time and resources to building dedicated infrastructure that supports their application processes. Implementing something similar in astronomy is a tall order and could negatively impact some perks of the individualized, largely open-admit model that our field has developed over time.

However, there are elements of each of these models that stand out in our research as changes that the GATF believes the AAS should recommend. These include:

1. Supporting a single central clearinghouse for letters of recommendation. This would substantially decrease workload for both students and letter writers, although adoption by programs would be more complex (see Section 6.2.1)
2. Recommending that departments adopt standardized application content (see Section 6.2.1).
3. Recommending a down-select date, where students admitted to a large number of programs would be strongly encouraged to limit their choices to just 2-3 schools by early April (see Section 6.2.2).

A centralized application system — one that includes some elements of both a common app and a matching process — is indeed one of the GATF's recommendations (see Section 6.4), with the caveat that it be implemented carefully and centrally over several years. This process will be complex and must be done with close consideration of its potential impacts on both applicants and programs, as well as how it aligns with potentially volatile and unpredictable budget and personnel realities. Our hope is that implementing the above steps will provide some of the benefits of such a system immediately and make an eventual large-scale implementation a smoother and less disruptive process.

# 5. Summary of Key Challenges

From our research a few key topics emerged as the biggest current challenges facing graduate admissions in astronomy.

Prospective graduate students find the application process to be overwhelming, expensive, and stressful. They find that the admissions process lacks transparency, with many individual issues stemming from poor communication by programs.

Graduate programs are grappling with unmanageable application numbers; many are seeing significant increases in their applicant pools. Programs are also struggling to manage and adapt their admissions processes in the midst of a rapidly-shifting funding and policy landscape.

Across the astronomy community there is frustration with the current process combined with a lack of clarity on how things can be improved. Common themes that emerged from our surveys, research, and conversations included a need for more information on how things are evolving and changing across the field, interest in some degree of standardization and centralization to help decrease workloads, and a desire for guidance from the AAS on useful resources and best practices to help improve all sides of the admissions process.

# 6. Recommendations and Next Steps

The GATF has four recommendations for actions the AAS could take to improve the current state of graduate admissions in astronomy. These include both directly addressing some of the key challenges presented above and setting up crucial infrastructure that will offer long-term support and resources to the astronomy community. The graduate admissions landscape is evolving rapidly, and making lasting positive changes will require both immediate action and longer-term in-depth work. Our recommendations focus on achieving both of these goals.

In brief, we recommend that the AAS:
- Establish a AAS Committee on the Status of Graduate Admissions in Astronomy (6.1)
- Recommend standardization of application content and key communication dates (6.2)
- Host a set of informational admissions webpages (6.3)
- Support adoption of a centralized "common application" system (6.4)

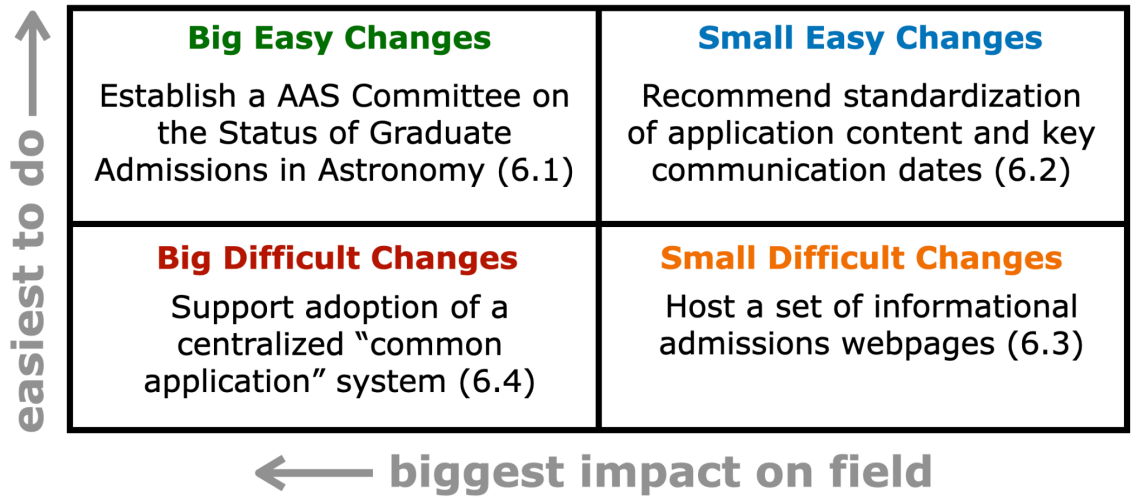

Impact-effort matrix for the GATF's four primary recommendations to the AAS.

We discuss each of these recommendations in more detail below.

## 6.1 Establish a AAS Committee on the Status of Graduate Admissions in Astronomy

Our first and strongest recommendation is the creation of a AAS Committee on the Status of Graduate Admissions in Astronomy.

It is clear that astronomy needs a centralized source of expertise and information when it comes to graduate admissions. Our admissions landscape is changing rapidly, and as a field we are navigating shared and complex challenges that require clear leadership and effective communication. A dedicated committee can serve as a common hub for communication, tracking changes in populations and policies throughout the field. In addition, our other recommendations include efforts that, if properly implemented, could encompass years of effort and require ongoing monitoring and dedicated follow-up, all tasks beyond the scope of a temporary task force and best suited for a permanent committee that can sustain communication with the field and build up institutional memory and expertise on astronomy graduate admissions.

Establishing a AAS committee focused on graduate admissions as one of the Society's Advisory Committees would also fill a unique role in the AAS. There is currently no AAS group tasked with overseeing the administration, best practices, and evolving policies that govern how students are admitted to graduate programs in astronomy. Many of these topics are unique to admissions itself and could not easily be fit under the umbrella of another committee.

There are, of course, complementary AAS committees that currently exist. The Education Committee oversees and provides advice on educational activities in the astronomy community, including undergraduate and graduate curricula and how these relate to student mentoring. The Committee on Employment is charged with facilitating professional development and employment at all career stages of astronomy. Finally, entities such as the Committee on the Status of Women in Astronomy (CSWA), the Committee on the Status of Minorities in Astronomy, (CSMA) and the Committee for Sexual-Orientation and Gender Minorities in Astronomy (SGMA) would all offer valuable insights on inclusive and equitable practices in our field. However, adding graduate admissions as a whole to any of these committees' portfolios would drastically expand their scope and dilute their other efforts.

A AAS "Admissions Committee" would function independently from these committees but have clear points of collaborative overlap with each, serving as both a focal point for communication on this multifaceted topic and a clear home for admissions-related tasks and responsibilities. Creating a AAS Admissions Committee will maximize the efficacy of the AAS as a whole at serving the needs of the community and overseeing the changes needed to address current challenges.

The single strongest example of this is the challenge of rapidly-increasing application numbers. It is clear from our applicant and program surveys that making the admissions process as streamlined and efficient as possible would enormously benefit applicants, advisers and letter writers, and admissions committees. This can be done through changes such as standardizing applications and making best-practices resources for applicants and committees more accessible. However, these changes will also make it easier for applicants to apply to more programs and exacerbate the problem of huge applicant pools.

There are three ways to address the challenge of large applicant pools, and all of them depend critically on the formation of a AAS committee focused on graduate admissions:

1. Our data point to the huge increase in the number of astronomy bachelors degrees (Section 1) as a key driver of growing applicant pools in astronomy. In reality, this is likely a combination of more people earning these bachelors degrees and a large fraction of them choosing to pursue PhDs. The growing number of people getting bachelors degrees in astronomy is under the purview of the AAS Education Committee. The many factors that drive degree holders to pursue graduate school as opposed to other careers is a topic best addressed by the AAS Employment Committee. The confluence of these phenomena, however, would be a direct focus of the AAS Committee on the Status of Graduate Admissions in Astronomy, which would serve as a crucial communication hub for addressing these topics.
2. We did not, in our data from recent years, see increases in the number of applications submitted per person (Section 2.2.1). However, if it becomes easier and more affordable to apply to multiple programs it is likely that this could change in the near future. This is not a change that could be easily addressed by a simple "best practices" recommendation or similar suggesting that students limit their application numbers: students who believe that more applications increase their odds of admission will have no incentive to comply. The only means of truly limiting application numbers is by adopting a centralized application system. While we recommend this in Section 6.4, we recognise that implementing this will be a lengthy and complex process that requires long-term effort and leadership at the AAS level. This can only be done through the dedicated efforts of a AAS Admissions Committee.
3. Finally, in the immediate future large applicant pools appear to be the simple reality of astronomy graduate admissions. With this in mind, a AAS Admissions Committee would be crucial for creating and maintaining the set of admissions webpages that we recommend in Section 6.3. These would include, for example, resources for admissions committees seeking to deal with large volumes of applicants while still maintaining fair and equitable admissions practices.

Beyond the immediate practical task of implementing our other recommendations and collaborating with other AAS committees, a AAS Admissions Committee would also fill a vital role for gathering and exchanging information across the field. This is particularly valuable in our current rapidly-changing admissions landscape. In the first months of 2025 a number of programs were forced to make rapid and drastic changes to their admissions process (including decreasing or even rescinding offers of admissions, modifying waitlist processes and policies, and altering offers of financial support) in response to substantial financial upheaval and uncertainty at the institutional, state, and federal levels.

In the crucial early days of these changes there was no single clear way for degree-granting programs to communicate with one another. This left many programs unclear on how their situations compared to others' in the field and was devastating for applicants who faced tremendous uncertainty in their application and financial status with little to no larger context. We expect such challenges to continue as the astronomy community adjusts to stark changes in how research and educational endeavors are supported in the United States: an [AIP survey](#) of both physics and astronomy graduate programs found that the number of first-year graduate students (in both physics and astronomy) was expected to decline by ~13% in the fall of 2025, with many programs noting that they expected larger cuts starting in the Fall of 2026. The creation of a AAS Admissions Committee would make it easier for the astronomy community as a whole to adjust to these changes in the area of graduate admissions, a process that is closely tied to federal funding, and is central to supporting the next generation of scientists.

An ideal AAS Committee on the Status of Graduate Admissions in Astronomy would be comprised of ~10-12 voting members, each with some experience in the admissions field and representing a range of career stages and degree-granting programs (including both MS and PhD programs and departments that are astronomy-specific or include astronomy alongside physics, planetary sciences, etc.) As with other advisory committees, members would be appointed by the Board of Trustees and serve a three-year term with the exception of the committee chair: committee chairs would serve for one year as a regular member, two years as chair (to prevent excessive leadership churn), and then spend their final year in a "past chair" role, for a total term of four years. Connecting with other related committees, such as the Education or Employment committees, will be an important part of the Admissions Committee's charter: this could be done at the membership level by including members that have current or past experience on such committees, or at the governance level through a shared Board of Trustees liaison.

We believe astronomy is facing unique and immediate challenges when it comes to graduate admissions and that establishing a AAS Committee on the Status of Graduate Admissions in Astronomy offers a suitably unique, doable, and effective solution.

## 6.2 Recommend standardization of application content and key communication dates

We propose that the AAS release an official recommendation for standardising key elements of the graduate admissions process. Specifically, the AAS should make recommendations regarding standardized application content and key communication points in the process between programs and applicants.

### 6.2.1 Recommendations for Standardized Application Content

Our research has confirmed that most astronomy graduate program applications have a shared DNA and large areas of overlap; however, the existing small variations in requirements and ambiguities in what programs are looking for currently make the applications process more difficult for students and the evaluation process more difficult for programs. Accordingly, we believe the AAS should release a recommendation endorsing some simple and straightforward standards for application content, including a letter submission platform, and a template for the application essay.

We want to stress that this would be a recommendation and nothing stronger. Some programs may be unable to adopt the standardized content due to restrictions from their universities, while others may simply be unwilling to due to a preference for their existing application. However, a recommendation that is broadly in line with what most programs are already doing will make adoption as easy as possible. This in turn will reduce the workloads on a) students that are applying to multiple programs, b) letter writers, and c) admissions committees that will be able to rely on a standard template (for the essay in particular) to help evaluate applications. An official statement from the AAS will offer programs a clear recommended template, which will help standardize any changes and give interested programs research-based backing when making the case to their home institutions.

Below we summarize what we find to be an effective recommendation for standard application content based on our survey results; however, this could be further refined by the AAS Admissions Committee:

**Curriculum Vitae, 1-2 pages:** conventionally this format is largely left up to students, but it is helpful to make clear that these are typically no more than 1-2 pages.

**Unofficial transcripts:** at the initial application review process unofficial transcripts should be requested if at all possible. Official transcripts are logistically and financially burdensome for applicants to request and, if need be, can be requested from admitted students at a later date.

**One 1500-word[1] application essay in a standard format:** a standard number (1) and length (1500 words, or approximately 3 pages) for essays will substantially reduce the burden on applicants to write and rewrite very similar content in a variety of different formations for each program. Even a minimal template for this essay would also help to standardise and clarify the content of the application: for example, an essay broken down into 1200 words encompassing the content currently contained in research and personal statements, followed by 300 words in which applicants clearly detail their interest in a specific program (put plainly, this serves as a formalization of the "tailoring paragraph" format that is already often adopted by applicants). A single widely-adopted format would make it easy for applicants to write a single main application and then modify it for different programs, and would also make it easy for admissions committees to equitably evaluate this written component of their applications by clarifying where applicants will address their preparedness, their personal background, and their interest in a program. This template could be as simple as the word-count breakdown described above; it could also take the form of a simple LaTeX/document template offered by the AAS (similar to the approach adopted by, for example, STScI telescope proposals).

**Two 500-word recommendation letters:** this decreases both the number of letters (two rather than the current typical standard of three) and their length (500 words corresponds to approximately one full page of single-space 12-point text, or about a page and a half of text on university letterhead).

In Section 4.2.3 we also note that a single central clearinghouse for these letters could be extremely beneficial for both recommenders and applicants. Letter writers would have a single location to submit a letter by a single effective due date (before their students' earliest application deadline). Students would only need to be concerned with seeing a single letter uploaded to a single location. This would also facilitate a shift to most letters being written as "general" rather than institutionally-tailored letters (as discussed in our "deep dive" conversations, reviewers often found program-tailored letters to be of little to no benefit, and in practice many letter writers are already pushed to do this; see Section 4.1).

At present, the main drawback to a central letter clearinghouse is a workload shift to administrators at individual institutions, many of whom would, at least in the immediate future, need to retrieve these letters from the central clearinghouse and add them to institution-specific platforms. This could prove to be prohibitively time intensive (imagine a department administrator downloading hundreds of letters and then uploading them one by one to a university website, taking care to match letters to the correct applicant). However, this should be a manageable problem provided that the central clearinghouse is chosen well with the goal of minimizing administrative overhead (for example, many REU programs use a Dropbox-style system or Google Forms, both of which have secure and user-friendly interfaces for retrieving uploaded documents or form entries). Programs could also consider the feasibility of "placeholder" documents to meet institutional requirements for a completed application without placing undue burdens on administrative staff.

---

[1] A word count rather than a page count is a much simpler and more straightforward way to establish and enforce length limits, as it avoids any ambiguity introduced by font choice, font size, margin size, line spacing, etc. and discourages applicants and readers alike from focusing on aesthetics rather than content.

The problem of the Likert-like scale rankings often associated with recommendation letters also remains. The GATF does not recommend including any such interface in the standardized application content since feedback on their usefulness was overwhelmingly negative; however, in practice these will likely remain a part of many programs' individual recommendation letter system, and in many cases these are a requirement set at a higher level (department, college, etc.) rather than by astronomy programs themselves. For the time being, we recommend that programs give clear guidance to admissions committee members on how data from these interfaces will be used. If letter writers are uploading their letters to program-specific websites, this same information should also be explicitly shared with letter writers; if letters are first uploaded to a central clearinghouse, this can be addressed at the administrative level.

**Optional/forbidden content:** we strongly recommend that committees not include additional application content described as optional. Committees should discourage applicants from including any additional information beyond the content requested in the official application; however, we also recommend that committees *not* explicitly forbid or ban inclusion of any information should an applicant wish to do so.

For example, we do not recommend that GRE or Physics GRE (PGRE) scores be included as part of the standardized astronomy graduate application. GRE and PGRE scores have both been shown to be poor predictors of students' future success in graduate school; the exams also propagate bias and introduce both financial and logistical barriers for many test-takers. In 2016 [the AAS recommended](#) that astronomy graduate programs "eliminate or make optional" the GRE and PGRE as an application requirement. In the years since, essentially all astronomy graduate programs and a large number of physics programs have dropped these exams, a change that was further accelerated during the COVID-19 pandemic when it became extremely difficult for students to take standardized tests. As of Apr 14, 2025, a crowdsourced document, "[GRE requirements & admissions fees for US/Canadian Astronomy & Physics Programs](#)", showed that 194 out of 199 of astronomy-related graduate programs in the US and Canada have dropped GRE/PGRE as an application requirement.

While the vast majority of programs have dropped the GRE/PGRE as a requirement (though a small number have recently re-introduced these exams or discussed the possibility in future years), one common approach (adopted by 61% of the schools in the "*GRE requirements*" list) has been to make it "optional" to some degree. Unfortunately, these sorts of optional application requirements can be badly misinterpreted by both applicants and admissions committees. In some cases, the inclusion of scores may simply be seen as a useful extra piece of information (with no explicit or implicit penalty applied to applicants who don't volunteer them), while in others, a lack of scores might be interpreted as a red flag or a de facto penalty (e.g. if Student A has strong PGRE scores while Student B reports none, an advantage is given to A). Most problematically, this interpretation could vary not only from program to program but from panelist to panelist within a single admissions committee. This is most easily avoided by transitioning this from an "optional" requirement (e.g. a statement that sending GRE/PGRE scores are optional, or including a non-required form field that prompts applicants to enter their scores[2]) to a simple statement that GRE scores will not be requested.

At the other extreme, *banning* even the mention of a GRE/PGRE score (in, for example, the application essay) is also a problem. First, any strong policy along these lines must come with the ability and willingness to enforce it, which takes administrative effort and can be inequitably applied. Second, in the specific case of the GRE/PGRE this can effectively serve to unfairly limit applicants who might feel, in some exceptional circumstance, that this information is important to the holistic case they are making in their application (the anecdotal example given is that of a student who, due to limited resources, has largely self-taught physics knowledge and wishes to demonstrate it through a test score in lieu of a transcript or recommendation letters).

Research portfolios have also been discussed as a tool for evaluating applications and student potential. Some schools (including one of our "deep dive" interviewees, as described in Section 4.1) use these very successfully and find them to be a valuable part of the application process. Portfolios were also suggested in the Barron et al. (2024) white paper on recommendation letters in astronomy. Unfortunately, it is not practical to make research portfolios a standard application requirement. Students' research products can vary extremely widely (first-author and co-author papers, conference proceedings or abstracts, posters, presentations, coursework). In practice the application essay and letters (following our suggested requirements above) already offer a clear way for applicants and advisers to summarize their research contributions. This is much more effective than placing the evaluation burden on admissions committees. We do not recommend research portfolios as part of the standard astronomy graduate application.

In lieu of requiring research portfolios, both making them optional and banning them outright also carry many of the same problems described for the GRE/PGRE: an optional requirement is prone to misinterpretation by both students and admissions committees, and a ban is difficult to enforce (particularly in edge cases such as a student including a link to a paper in their CV or essay).

---

[2] We recognize that some programs do not use the GRE and/or PGRE in their evaluations but nevertheless have an application form that requests them due to institutional requirements that cannot easily be changed.

Rather than dealing with optional or forbidden content, programs should simply make clear, to both applicants and admissions committees, what the application requirements are and how any unrequested content will be used.

**Application fees:** we recommend eliminating or reducing application fees whenever possible, as these fees are often a significant burden on applicants (see Section 2.2.1) and – depending on how finances vary between institutions – offer little to no benefit to admissions committees beyond serving as an artificial and highly inequitable financial barrier. However, since the administration of application fees is often beyond the control of individual programs, we recognize that this may not be possible in many cases. In such scenarios, we instead recommend that programs make any available fee waivers or other other financial aid processes as accessible and transparent as possible, with the goal of removing financial burden as a barrier for prospective applicants.

### 6.2.2 Recommendations for Timelines and Communication

The simplest and most important recommendation that the AAS can make regarding timelines is to encourage transparency and clarity from programs when communicating dates with applicants. Our applicant survey highlighted poor communication and opaque program information as a significant source of stress in the application process, and this uncertainty can lead to delays in the process and place enormous time pressure on programs and applicants alike to make critical decisions about offers and acceptances. Many applicants noted that "Lack of process transparency" was especially stressful for them, and some noted specifically that the ambiguity on when they would hear back from the program was especially difficult. Put simply, the AAS should encourage graduate programs to clearly state and commit to key dates when they will communicate with applicants.

A clearly-publicized "notification date" from a program, for example, would inform applicants that they can expect concrete news on their application — admission, rejection, or placement on a waitlist — by a specific date (it is, of course, crucial that programs are able to then follow through on this commitment). This would give applicants a clearer and less stressful timeline for when they can expect to hear back, and when they will have a complete set of information regarding their grad school choices and can proceed to the decision-making stage. Programs with other key interim dates — for example, contacting longest applicants to arrange interferes — could make these similarly clear.

Standardizing the dates themselves is a much more challenging proposition, simply because programs often face external time pressures that are difficult to shift. Our "deep dive" conversations confirmed that universities often have unshiftable dates in place when it comes to, for example, school-wide application deadlines, application fee waivers, or nominations for internal fellowships that require offers be made by a certain date. Other programs prefer an earlier or later timeline, particularly terminal Masters or Bridge programs that intentionally operate at an offset to the PhD admissions process. Other important dates, including holidays and the announcement of external fellowships such as the NSF GRFP, vary from year to year or, in the latter case, aren't shared ahead of time. Finally, programs generally *prefer* to stagger things such as interview dates or visit dates to best accommodate applicants' schedules.

April 15th is a (partial) exception to this standardization challenge. This comes from an agreement by the Council of Graduate Schools known as the "April 15th Resolution", and it specifically states that "[s]tudents are under no obligation to respond to offers of financial support prior to April 15; earlier deadlines for acceptance of such offers violate the intent of this Resolution." However, it's important to note that this resolution does not actually *set* April 15th as a deadline; it merely prohibits schools from demanding decisions earlier than April 15th. That said, in reality many graduate schools have set April 15th as their official deadline thanks to this Resolution. Others set it as a de facto deadline, but retain the ability to make occasional offers or accept decisions after April 15th. The result is that, while April 15th looms large as a decision date for graduate school, the particulars of how it is implemented still vary from program to program.

Based on the April 15th decision date, and following from our research presented in Section 4.2, one field-wide date that would be relatively easy to both recommend and adopt would be a "down-select" or "recommended reduction" date, when applicants with multiple offers are strongly encouraged to narrow their choices down to their top two or three choices. Our applicant survey revealed that 45% of applicants who received offers receive them from 3 or more schools; while this is not a majority of applicants, it nevertheless represents a significant fraction of "slots" that could be freed up if applicants are encouraged to narrow down their final choice to two programs and decline offers ahead of the April 15th deadline. Because of the April 15th Resolution, this would need to function as a recommendation rather than an enforceable date, but even this could help alleviate the problem of last-minute decisions in the final days of the admissions process.

Knowing that full timeline standardization is unlikely, an "example" timeline based on our survey results and refined by the AAS Admissions Committee, could prove informative for all programs and offer a useful template for those that have even partial flexibility or interest in standardizing. This could resemble the following timeline, a combination of the key/most common existing dates in astronomy grad admissions and our proposed recommendations:

**Application deadline (Dec 1st):** deadline for submitting applications and recommendation letters
**First application update (Jan 15th):** applicants are informed on this date if they are rejected, waitlisted, or being invited to interview
**Second application update (Feb 1st):** remaining applicants are informed on this date if they are rejected, waitlisted, or receiving an offer of admission
(note that this timeline leaves much of February and March open for flexibility in scheduling virtual or in-person visits and allowing applicants time to make decisions)

**Down-select date (Apr 1st):** applicants with multiple offers are strongly encouraged to narrow their choices down to their top two and officially decline other offers.
**Decision date (April 15th):** official decision deadline

Several key dates and decision points are still left open to interpretation by this timeline; this is meant to be a fairly minimalist guide. For example, whether and when to schedule interviews (typically done in January after several rounds of application review and the winter AAS), arranging virtual or in-person visits (typically done in February and March), and the hard-to-standardize process of making offers off the waitlist (typically done in March or April) are all left up to individual programs. Still, even this fairly minimalist guide includes critical communication and decision dates that, if broadly adopted, would significantly streamline the graduate admissions process.

### 6.2.3 Expected Impact

This is a lengthy section, and it may make this recommendation seem overwhelming or extreme. However, the goal of the GATF here is to apply a light touch and (with the exception of the reduction from 3 to 2 recommendation letters and the "down-select date") propose a format and timeline that is very close to what most programs are already doing: programs are likely already following some or even all of the requirements and practices described above. Our goal is to move toward large-scale agreement that requires minimal changes on the program side while significantly benefitting applicants.

Consider, for example, an applicant writing ten essays where five are required to be 1500-words and the other five are required to be three pages; two of these are due November 30, five are due December 1, and three are due on December 2nd. In practice these requirements are essentially identical, but the student may find themselves double-checking each individual program's stated requirement and deadline, fretting over word- or sentence-scale edits, and putting in time-consuming effort for changes that have little to no actual impact on their applications' content and strength. By contrast, if all of these programs shifted to a 1500-word requirement and a December 1st due date, it would alleviate a significant burden on the applicant while having next to no tangible impact on programs' timelines or the length and quality of the applications they receive and read.

However, the biggest benefit to standardizing application content – making the process easier and more streamlined for students – carries with it a significant potential challenge: this could lead to a marked increase in the number of applications that each student submits. Students cite the current process (e.g. writing application essays, tailoring these essays by program, and keeping track of multiple deadlines) as a burdensome ordeal that limits their total number of submitted applications (and recall that our applicant survey preferentially sampled "successful" applicants who ultimately completed the process and were admitted to grad school; see Section 2.3). It's important to keep in mind that limiting the number of submitted graduate school applications based on students' time, money, and resources is not an effective or equitable way to control numbers since it risks discouraging promising students that cannot currently meet the demands of a needlessly inefficient process. It's also worth remembering that some programs remain eager for increases in their applicant pools. Even so, an increase in applications if this recommendation is adopted could further strain programs that are already grappling with unmanageable applicant pools.

Some of these effects can be mitigated by the standardization process itself: decreasing the number of length of recommendation letters, limiting applications to a single three-page essay, and adhering to even a minimal template for essays should make the reading and reviewing process easier for admissions committees. Timeline recommendations can also help keep workloads manageable, particularly on the decision end of the process. However, fully dealing with already-large and increasing application numbers, on both the applicant and admissions committee sides, demands further changes. Some are addressed above in our discussion of the role of the AAS Committee on the Status of Graduate Admissions in Astronomy (Section 6.2.1); others are presented below.

## 6.3 Host a set of informational admissions webpages

We recommend that the AAS create and host a set of graduate admissions webpages on aas.org, containing key information for community members participating in various aspects of the admissions process. These pages would serve as a broadly accessible, easy-to-find, and well-maintained community resource.

The technical task of creating these informational pages should be straightforward, requiring minimal staff hours and support to add them to the existing AAS website infrastructure. However, curating and compiling the information on the pages will be a lengthier job and one that would be an important early priority of the AAS Admissions Committee (Section 6.1). Below we list a few suggested examples for the sorts of information that we believe would be most valuable as online community resources.

**Practical program information:** There is currently no official AAS list of programs that offer graduate degrees in astronomy. Students often learn of programs via word-of-mouth (including speaking with professors, advisers, peers, and representatives at events like the annual Graduate School fair at the winter AAS meeting) or their own research. The Division of Planetary Sciences maintains a [list of programs](), both in the US and Canada and international, that "can lead to a PhD with a planetary science focus" (in April 2025 it included 95 programs), along with a [spreadsheet]() that lists adviser-specific open MS/PhD positions in planetary science for the coming year (in April 2025 it listed 46 positions starting in the fall of 2025). Recently, the [crowdsourced GRE policy document]() came to serve as a de facto list of potential programs and useful application information for prospective students, listing 199 programs in April of 2025. The American Institute of Physics also maintains the [GradschoolShopper.com]() website as a resource for researching graduate programs in the physical sciences; a search for "astronomy" in April 2025 returned 198 results.

Even a simple text list on the AAS website, or a summary of the pages and resources listed above, would be a valuable guide for prospective students and advisers alike. Additional useful information could include direct links to programs' current admissions websites and contact information, or even (following the general model of [GradschoolShopper.com](GradschoolShopper.com)) links to short summary pages for each program that list yearly up-to-date information on deadlines, fee waivers, admissions policies, and other key details.

**Information for applicants and mentors:** a page aimed at applicants could include information such as examples CVs and essays, best-practices guides for things like soliciting letters of recommendation and contacting prospective advisers, and summaries of some basic statistics on astronomy graduate admissions (including some of the data presented in this report).

Mentors and advisers are a key part of the graduate admissions process on the applicant side; the data and best-practices suggestions described above will also be a useful resource for those advising undergraduate students. In addition, a webpage aimed specifically at advisers could include a compilation of crucial information on writing letters of recommendation. There is a significant body of research on how to write effective reference letters and avoid perpetuating bias and inequities. Sharing a few key recommended links would be an extremely valuable resource for letter writers.

**Information for admissions committees and programs:** a AAS webpage aimed at admissions committee chairs and members could include information on recommended application content (see Section 6.2.1) and communication dates (Section 6.2.2). This would also be an ideal place to share guidelines on things like evaluation rubrics and best practices for conducting and assessing interviews (both cited by programs as areas where clear guidance from the AAS would be helpful; see Section 4.1); much like reference letters, there is ample information online suggesting how to use rubrics and conduct admissions interviews, so a AAS webpage that serves primarily to compile a select subset of current research would be a useful way of helping committees to access the information they need. This page could also be used to publicly share up-to-date data on the astronomy-wide graduate admissions landscape, including results from this report and other current or future surveys of graduate programs evaluating, for example, the ongoing impact of federal funding and policy shifts.

Finally, an online AAS presence supporting graduate admissions could also be a useful means of hosting live or recorded webinars, workshops, panel discussions, Q&A sessions, or other multimodal resources aimed at all of the groups listed above.

Crucially, all of these webpages will only be useful if the content is kept up-to-date and carefully curated. Maintaining a page with current program information will require soliciting yearly updates from programs, a task that would ideally be undertaken by the AAS Admissions

Committee. Similarly, identifying the most valuable research summaries and resources for things like writing letters, using rubrics, and conducting interviews is work that should also be done by the AAS Admissions Committee in collaboration with other AAS committees (Education, Employment, CSWA, CSMA, SGMA) that could offer relevant expertise.

## 6.4 Support adoption of a centralized "common application" system

We recommend that the AAS support the adoption of a centralized "common app" system for astronomy graduate admissions.

The GATF believes that a well-executed central application system for graduate admissions would be extremely beneficial to the astronomy community. Done well, such a system would decrease both applicant and committee workloads and improve outcomes for both students and programs. However, this would be a substantial and complicated change to how our field approaches graduate admissions. Designing, deploying, and supporting a centralized system requires a long-term investment of time and effort.

Our first three recommendations — creating a AAS Committee on the Status of Graduate Admissions in Astronomy, supporting standardized application content and key communication dates, and hosting informational webpages — all lay valuable groundwork for this process. The AAS Admissions Committee will be the primary hub for researching the implementation of a centralized application system, assessing its impact, and gathering feedback from the community. If programs adopt AAS-recommended application content and communication dates within their own admissions processes, this will help ease the eventual transition to a centralized system. Graduate admissions webpages hosted by the AAS will quickly establish a central resource for information and updates.

A detailed roadmap for implementing a centralized application system is beyond the scope of our relatively short-lived task force. We can, however, highlight a few important challenges that should be addressed, along with potential solutions.

**Application tailoring:** In Section 6.2 we include suggestions for standardizing the content of applications; one component of this is a 1500-word application essay that includes a dedicated section where applicants can discuss their interest in a particular program. If astronomy shifted to a straightforward "common-app" system, students would, in principle, submit a single application, including a single essay that would not be tailored to their chosen programs.

A more effective common-app system is one that allows a limited amount of tailoring. Students could, for example, submit a standard 1200-word essay to the system along with tailored 300-word write-ups for each program where they plan to apply. Programs could, if desired, provide their own prompts for these write-ups. However, this approach should be implemented carefully, since excessive tailoring options will ultimately dilute the workload-reducing benefits of a "common-app" process.

**Program adoption barriers:** some astronomy graduate programs might be eager to join a centralized common application system and able to do so easily. However, others might face complications ranging from college or university restrictions (such as internal deadlines or requirements that graduate applications be submitted through their own system) to pushback in departments where astronomy shares some or all of its admissions administrative burden with other fields such as physics or planetary science. Moving to a centralized system without considering these drawbacks could place a substantial administrative burden on programs' staff members or lead to logistical issues (such as conflicts with internal deadlines) that make their admissions processes less effective or efficient.

Some of these challenges can be addressed within programs by discussions among faculty and staff. Others can be addressed through communication with the AAS Admissions Committee. However, it's also important to determine how widely-adopted a central application system would need to be in order to be considered effective. Undergraduate common-app systems, for example, are not universally used, but even participation by a subset of programs makes them useful tools. Defining this threshold, and understanding how to achieve it, is an important early question to answer when implementing a centralized application system.

**Managing application numbers:** as noted above, standardization of application content could lead to an increase in the number of applications submitted per person, particularly if this is eventually combined with an easy-to-use and accessible central submissions system.

One simple way of addressing this is to directly limit how many applications students can submit. This could function as simply as students submitting a single application and then selecting which programs should receive it, with the system limiting the number of selections: a limit of 10-12 would be in good agreement with the average number of programs that students have been applying to over recent cycles. However, this approach has notable potential drawbacks. Students' selections would almost certainly not be evenly distributed: "popular" programs could continue to receive unmanageably large numbers of applications while smaller or less-well-known programs could even risk seeing a drop in their application numbers.

Taking applicants' interest level in programs into consideration could be another way of allowing programs to control the number of applications they review while also potentially increasing yield. In this model, students could rank the programs that they apply to or parse them into several broader size-limited categories (for example, designating programs where they are "very interested", "interested", and "potentially interested" in attending). Programs could then use these rankings as a means of limiting how many applications they review: some programs may only wish to evaluate applicants who are "very interested" in attending, while others may prefer to see all interested applicants. This could help programs that receive a large number of applications to limit their review pool to students who have a relatively high chance of accepting an offer, while programs that are still seeking to grow their applicant pool can review more broadly. These rankings would almost certainly need to be non-binding (unlike the strict rankings used by the medical residency matching process); allowing prospective students to learn more about potential programs and reevaluate their preferences during the decision process is a key component of the current admissions process in astronomy. However, non-binding rankings could also make it possible for applicants to attempt to rank programs strategically rather than in a way that truly reflects their interest level.

Another approach worth exploring, particularly in the current landscape of large applicant pools and small and uncertain program sizes, is a variation on the SOAP or "Scramble" component of the medical residency matching process (see Section 4.2). The goal of this approach would be to effectively connect applicants looking for programs and programs looking for applicants in the final stages of the admissions process. Near the end of the process (just after, for example, the "down-select" date described in Section 6.2.2, or after the earliest decision deadline of April 15th) programs still in search of students could identify themselves. Applicants who had not previously applied and who were still in search of a graduate program could then opt into sending these programs their applications. Ideally, this would help to minimize the dual problems of strong applicants who receive no offers and programs that wish to recruit more students.

The questions of how to retain the most successful elements of astronomy's current approach to graduate admissions, avoid extreme and unintended changes in applicant pool size, and encourage ease of adoption by graduate programs are all concerns that must be addressed when developing a centralized application system. Success will require exploring all of the approaches described above and likely adopting a combination of these and other solutions However, if implemented properly, a centralized system could significantly decrease workloads across the entire community, support a clear, easy-to-use, and transparent admissions process, and increase both success rates for applicants and yield rates for programs. This should be one of the main goals and focus areas of the AAS Committee on the Status of Graduate Admissions in Astronomy.

# Appendix

For reference, PDF copies of the questions asked as part of our Applicant Survey and Program Survey are included below.

# AAS Graduate Admissions Feedback Form

The purpose of this form is to collect information from **recent applicants** on their experiences and concerns regarding the graduate admissions process. This survey is geared towards any person who has applied to Astronomy or Astrophysics PhD serving institutions in the last ~5 years regardless of their current career path or academic standing.

This form will also serve as a place to provide input on how the astronomy community might improve the current state of affairs.

* Indicates required question

1. Have you applied to Astronomy or Astrophysics related graduate programs in the US within the last ~5 years? *

   *Mark only one oval.*

   ◯ Yes    *Skip to question 3*
   ◯ No

2. Select which type of US-based, Astronomy or Astrophysics related graduate programs you have ever applied to within the last 5 years? *

   *Check all that apply.*

   ☐ PhD
   ☐ Masters
   ☐ Other: _______________

ACADEMIC INFORMATION

3. **What is your current academic status? ***

   *Mark only one oval.*

   ◯ Undergraduate Student
   ◯ Post-Baccalaureate Student
   ◯ Masters Student
   ◯ PhD Student/Candidate
   ◯ Post-Doctoral Fellow
   ◯ Non-academic Position inside the field of Astronomy
   ◯ Academic Position outside the field of Astronomy
   ◯ Non-academic Position outside the field of Astronomy
   ◯ Other: _______________________

4. **What is the highest level of education you have completed to date? ***

   *Mark only one oval.*

   ◯ High school graduate, diploma or equivalent (e.g., GED)
   ◯ Some college, no degree
   ◯ Associate degree
   ◯ Bachelor's degree
   ◯ Master's degree
   ◯ Doctorate degree
   ◯ Other: _______________________

5. **What kind of university did you attend for your undergraduate studies? (check all that apply) ***

   *Check all that apply.*

   ☐ US Private
   ☐ US Public
   ☐ International Private
   ☐ International Public

6. **For any universities attended in the United States, please specify their designations (check all that apply)** *

   *Check all that apply.*
   - [ ] R1 Research University "Very High Research Activity"
   - [ ] R2 Research University "High Research Activity"
   - [ ] R3 Research University "Moderate Research Activity"
   - [ ] Liberal-Arts College
   - [ ] Minority-Serving Institutions (MSI)
   - [ ] Historically Women's Colleges
   - [ ] Community College
   - [ ] Other: _______________

7. **What kind of degree did/will you receive from your undergraduate institution(s)? (check all that apply)** *

   *Check all that apply.*
   - [ ] B.S./B.A./B.S.C in Astronomy/Astrophysics
   - [ ] B.S./B.A./B.S.C in Physics
   - [ ] B.S./B.A./B.S.C in Earth and Planetary Sciences
   - [ ] B.S./B.A./B.S.C/B.tech in Engineering
   - [ ] Other: _______________

Undergraduate Mentoring Programs 1

8. **Did you participate in any mentoring programs/workshops specifically designed to help students apply to graduate school?** *

   *Mark only one oval.*
   - ( ) Yes   *Skip to question 9*
   - ( ) No    *Skip to question 12*

Undergraduate Mentoring Programs 2

9. **Was your undergraduate mentoring program affiliated with NSF's Research Experiences for Undergraduates (REU)** *

    *Mark only one oval.*

    ○ Yes
    ○ No
    ○ Unsure

10. **Did you have to apply to this mentoring program?** *

    *Mark only one oval.*

    ○ Yes
    ○ No

11. **How did your undergraduate mentoring program help you prepare for the application process (check all that apply)** *

    *Check all that apply.*

    ☐ FAQ sessions
    ☐ Interview prep
    ☐ Essay workshopping
    ☐ Securing letters of recommendation
    ☐ Other: _______________

Undergrad Research 1

12. **Did you do research in the field of Astronomy or Physics as an undergraduate student?** *

    *Mark only one oval.*

    ○ Yes     *Skip to question 13*
    ○ No      *Skip to question 19*

Undergraduate Research 2

13. **When did you start undergraduate research?** *

    *Mark only one oval.*

    - ◯ 1st year
    - ◯ 2nd year
    - ◯ 3rd year
    - ◯ 4th year
    - ◯ Other: ___________

14. **How many different projects did you work on as an undergraduate researcher?** *

    *Mark only one oval.*

    - ◯ 1 project
    - ◯ 2 projects
    - ◯ 3 or more projects

15. **Did you participate in REU programs as an undergraduate researcher? If so, how many?** *

    *Mark only one oval.*

    - ◯ No, I did not participate in an REU program
    - ◯ 1 program
    - ◯ 2 programs
    - ◯ 3+ programs

16. **Did your undergraduate research projects result in any first-author publications that were submitted by the time you applied to graduate programs?** *

    If so, how many did you have <u>by the time you applied to graduate programs</u>?

    *Mark only one oval.*

    - ◯ 0 publications
    - ◯ 1 publication
    - ◯ 2 publications
    - ◯ > 3 publications

17. Did your undergraduate research projects result in any submitted **co-author** publications, by the time you applied to graduate programs?

    If so, how many did you have <u>by the time you applied to graduate programs</u>?

    *Mark only one oval.*

    - ◯ 0 publications
    - ◯ 1 publication
    - ◯ 2 publications
    - ◯ 3+ publications

18. By the time you applied to graduate programs, had you given any professional research presentations?
    i.e.: Poster or talk at AAS or another conference, undergraduate symposiums, etc

    *Mark only one oval.*

    - ◯ Yes
    - ◯ No

   APPLICATION PROCESS

19. How long has it been since you last applied to an Astronomy graduate program? *

    *Mark only one oval.*

    - ◯ < 1 year
    - ◯ < 2 years
    - ◯ < 3 years
    - ◯ < 4 years

20. How many schools/programs did you apply for in your most recent application cycle?

    _______________________________

21. During which school year did you most recently apply?    Dropdown

    Mark only one oval.

    - Fall 2023
    - Fall 2022
    - Fall 2021
    - Fall 2020
    - Fall 2019
    - Fall 2018
    - before Fall 2018

22. How many fee waivers did you apply for in your most recent cycle?

    ___________________________________________

23. Did the application fees limit the number of programs you applied to?

    Mark only one oval.

    - Yes
    - No
    - Maybe

24. Were the application fees financially burdensome for you?

    Mark only one oval.

    - Yes
    - No
    - Somewhat

25. Would you have applied to more schools if there was a standardized platform, such as the Common App?

    *Mark only one oval.*

    ◯ Yes

    ◯ No

    ◯ Maybe

26. How many of these schools/programs did you interview for? *

    ______________________________

27. How many of the schools/programs you applied to extended you an offer? *

    ______________________________

28. How many of the schools/programs you applied to waitlisted your application? *

    ______________________________

29. How many of the schools/programs you applied to rejected your application? *

    ______________________________

30. In total, how many admissions cycles have you applied for? *

    *Mark only one oval.*

    ◯ 1      Skip to question 37

    ◯ 2      Skip to question 31

    ◯ 3+     Skip to question 31

**Next we'd like to ask you about the application cycles prior to the most recent one you answered about previously.**

31. For any previous cycle during which you applied to programs, please indicate the year of each
   (A: Most recent - C: Least recent)

   Mark only one oval per row.

   |  | 2022 | 2021 | 2020 | 2019 | 2018 | N/A |
   |---|---|---|---|---|---|---|
   | Cycle A | ○ | ○ | ○ | ○ | ○ | ○ |
   | Cycle B | ○ | ○ | ○ | ○ | ○ | ○ |
   | Cycle C | ○ | ○ | ○ | ○ | ○ | ○ |

32. For each cycle you listed above, please indicate how many schools/programs **you applied for**?

   Mark only one oval per row.

   |  | Fewer than 5 | 5 - 10 | 10 - 15 | 15 - 20 | More than 20 | N/A |
   |---|---|---|---|---|---|---|
   | Cycle A | ○ | ○ | ○ | ○ | ○ | ○ |
   | Cycle B | ○ | ○ | ○ | ○ | ○ | ○ |
   | Cycle C | ○ | ○ | ○ | ○ | ○ | ○ |

   How many of these schools/programs extended you an offer?

33. For each cycle you listed above, please indicate how many schools/programs **extended you an offer**?

    *Mark only one oval per row.*

    |  | Fewer than 5 | 5 - 10 | 10 - 15 | 15 - 20 | More than 20 | N/A |
    |---|---|---|---|---|---|---|
    | Cycle A | ○ | ○ | ○ | ○ | ○ | ○ |
    | Cycle B | ○ | ○ | ○ | ○ | ○ | ○ |
    | Cycle C | ○ | ○ | ○ | ○ | ○ | ○ |

    How many of these schools/programs did you interview for?

34. For each cycle you listed above, please indicate how many schools/programs **you interviewed for**?

    *Mark only one oval per row.*

    |  | Fewer than 5 | 5 - 10 | 10 - 15 | 15 - 20 | More than 20 | N/A |
    |---|---|---|---|---|---|---|
    | Cycle A | ○ | ○ | ○ | ○ | ○ | ○ |
    | Cycle B | ○ | ○ | ○ | ○ | ○ | ○ |
    | Cycle C | ○ | ○ | ○ | ○ | ○ | ○ |

    How many of these schools/programs waitlisted your application?

35. For each cycle you listed above, please indicate how many schools/programs **waitlisted your application**?

    *Mark only one oval per row.*

    |  | Fewer than 5 | 5 - 10 | 10 - 15 | 15 - 20 | More than 20 | N/A |
    |---|---|---|---|---|---|---|
    | Cycle A | ○ | ○ | ○ | ○ | ○ | ○ |
    | Cycle B | ○ | ○ | ○ | ○ | ○ | ○ |
    | Cycle C | ○ | ○ | ○ | ○ | ○ | ○ |

    How many of these schools/programs rejected your application?

36. For each cycle you listed above, please indicate how many schools/programs **rejected your application**?

    *Mark only one oval per row.*

    |  | Fewer than 5 | 5 - 10 | 10 - 15 | 15 - 20 | More than 20 | N/A |
    |---|---|---|---|---|---|---|
    | Cycle A | ○ | ○ | ○ | ○ | ○ | ○ |
    | Cycle B | ○ | ○ | ○ | ○ | ○ | ○ |
    | Cycle C | ○ | ○ | ○ | ○ | ○ | ○ |

## Application Decisions

Thinking about your most recent application experiences, what factors impacted your decisions?

37. How important were these factors in <u>deciding where to apply?</u>

*Mark only one oval per row.*

| | Not Important | Slightly Important | Important | Very Important |
|---|---|---|---|---|
| GRE/PGRE requirements | ○ | ○ | ○ | ○ |
| Local socio-political climate | ○ | ○ | ○ | ○ |
| Physical location | ○ | ○ | ○ | ○ |
| Program prestige | ○ | ○ | ○ | ○ |
| Faculty research | ○ | ○ | ○ | ○ |
| Telescope access | ○ | ○ | ○ | ○ |
| Initial funding availability | ○ | ○ | ○ | ○ |
| Long-term funding stability | ○ | ○ | ○ | ○ |
| Union/Unionization Efforts | ○ | ○ | ○ | ○ |
| Cost of Living | ○ | ○ | ○ | ○ |
| Typical time to degree | ○ | ○ | ○ | ○ |
| Departmental culture and values | ○ | ○ | ○ | ○ |

38. If you accepted an offer, how important were these factors in your final decision? Please note the rows are different in this question.

*Mark only one oval per row.*

| | Not Important | Slightly Important | Important | Very Important |
|---|---|---|---|---|
| Local socio-political climate | ○ | ○ | ○ | ○ |
| Physical location | ○ | ○ | ○ | ○ |
| Program prestige | ○ | ○ | ○ | ○ |
| Faculty research | ○ | ○ | ○ | ○ |
| Telescope access | ○ | ○ | ○ | ○ |
| Initial funding availability | ○ | ○ | ○ | ○ |
| Long-term funding stability | ○ | ○ | ○ | ○ |
| Union/Unionization Efforts | ○ | ○ | ○ | ○ |
| Cost of Living | ○ | ○ | ○ | ○ |
| Typical time to degree | ○ | ○ | ○ | ○ |
| Departmental culture and values | ○ | ○ | ○ | ○ |

39. Were there any factors not listed above that affected your decisions on where you applied/accepted an offer from? If so, how important were those factors?

______________________________________

40. If you visited any of the schools, virtually or in-person, how much did the visit(s) impact your final decision? *

    *Mark only one oval.*

    ◯ Did not attend a grad school visit
    ◯ Significantly impacted the decision
    ◯ Slightly impacted the decision
    ◯ Did not impact the decision

41. Did you apply for graduate fellowships (NSF GRFP, Ford, DOE, etc.) as an undergraduate ? *

    *Mark only one oval.*

    ◯ Yes     *Skip to question 42*
    ◯ No      *Skip to question 46*

Graduate Fellowships

42. How many graduate fellowships (NSF GRFP, Ford, DOE, etc.) did you apply for? *

    *Mark only one oval.*

    ◯ 1
    ◯ 2
    ◯ > 3

43. Were you awarded any graduate fellowships (NSF GRFP, Ford, DOE, etc.) ? *

    *Mark only one oval.*

    ◯ Yes     *Skip to question 44*
    ◯ No      *Skip to question 46*

Successful Graduate Fellowships

44. Did your fellowship impact where you chose to go to grad school? *

    *Mark only one oval.*

    ◯ Yes
    ◯ No
    ◯ Other: _____________

45. Did your fellowship impact when you accepted an admissions offer? *

    *Mark only one oval.*

    ◯ Yes, I waited until the fellowships were announced
    ◯ No, I did not wait for a fellowship before accepting a grad school offer
    ◯ Other: _____________

## Full Demographic Information (OPTIONAL)

This section asks about demographic information and is **optional.** This survey is entirely anonymous and these responses will not be published in case-by-case manner.

46. What is your age?

    _____________

47. What is your race/ethnicity? (Select all that apply)

    *Check all that apply.*

    ☐ White/Caucasian
    ☐ Black/African American
    ☐ Hispanic/Latino
    ☐ Asian
    ☐ Native American/Alaska Native
    ☐ Native Hawaiian/Other Pacific Islander
    ☐ Middle Eastern/North African
    ☐ Multiracial
    ☐ Other: _____________

48. What is your primary language? What language did you speak growing up?

    *Mark only one oval.*

    ◯ English
    ◯ Spanish
    ◯ Mandarin or Cantonese
    ◯ Hindi or Bengali
    ◯ Other: _________________________

49. Are you the first in your family (excluding siblings) to attempt a Bachelor's degree?

    *Mark only one oval.*

    ◯ Yes
    ◯ No

50. Are you the first in your family (excluding siblings) to attempt a graduate-level degree?

    *Mark only one oval.*

    ◯ Yes
    ◯ No

51. What is your gender identity? (select all that apply)

    *Check all that apply.*

    ☐ Man
    ☐ Woman
    ☐ Non-Binary
    ☐ Transgender
    ☐ Agender/I don't identify with any gender
    ☐ Other: _________________________

52. **What is your sexual orientation?**

    *Mark only one oval.*

    ○ Heterosexual
    ○ Homosexual
    ○ Bisexual
    ○ Pansexual
    ○ Asexual
    ○ Queer
    ○ Other: _______________________

## Open Feedback

53. **What did you find to be the most challenging part of the graduate admissions process?**

    _________________________________________________
    _________________________________________________
    _________________________________________________
    _________________________________________________
    _________________________________________________

54. **If you had to pick one thing to change regarding graduate admissions process, what would it be?**

    _________________________________________________
    _________________________________________________
    _________________________________________________
    _________________________________________________
    _________________________________________________

55. **Do you feel your academic trajectory was unique in anyway? Please elaborate.**

    _________________________________________________
    _________________________________________________
    _________________________________________________
    _________________________________________________
    _________________________________________________

56. Is there anything else you would like to share with the AAS regarding the graduate admissions process?

# AAS Graduate Admissions Department Survey

## Introduction

**The purpose of this survey is to collect information from graduate programs that award degrees in astronomy, astrophysics, or related fields (e.g. physics) where the research focus is astronomy- or astrophysics-related. The American Astronomical Society (AAS) is interested in collecting data on recent trends in applications and on application procedures and policies from these programs. This form will also serve as a place to provide input on how the astronomy community might improve the current state of affairs in graduate admissions.**

# AAS Graduate Admissions Department Survey

## Program Details

* 1. What are the designations of your institution (check all that apply)?
  - [ ] R1 Research University ("Very High Research Activity")
  - [ ] R2 Research University ("High Research Activity")
  - [ ] R3 Research University ("Moderate Research Activity")
  - [ ] Liberal Arts College
  - [ ] Public Institution
  - [ ] Private Institution

* 2. What type of department/program are you?
  - [ ] Astronomy or Astrophysics Department
  - [ ] Physics and Astronomy Department
  - [ ] Physics Department
  - [ ] Planetary Science Department
  - [ ] Other (please specify)

# AAS Graduate Admissions Department Survey

## PhD Applications

* 3. Do you accept applications for the PhD?

| AAS Graduate Admissions Department Survey |
|---|
| PhD Application Numbers and Timing |

**Please note that we request information for the past 6 years in order to get information on trends that include one full year before the COVID-19 pandemic.**

4. How many PhD applications did you receive in the following academic years?

2023-2024: [ ]
2022-2023: [ ]
2021-2022: [ ]
2020-2021: [ ]
2019-2020: [ ]
2018-2019: [ ]

5. How many PhD offers did you make in each of these years?

2023-2024: [ ]
2022-2023: [ ]
2021-2022: [ ]
2020-2021: [ ]
2019-2020: [ ]
2018-2019: [ ]

6. How many PhD offers were accepted in each of these years?

2023-2024: [ ]
2022-2023: [ ]
2021-2022: [ ]
2020-2021: [ ]
2019-2020: [ ]
2018-2019: [ ]

7. When was your most recent admissions application deadline (mm/dd/yy)?

[ ]

8. When did you make your first admissions offers this year?

   ○ Jan 1-15
   ○ Jan 16-31
   ○ Feb 1-15
   ○ Feb 16-29
   ○ Mar/Apr
   ○ Other (please specify)

9. When was your standard acceptance deadline (mm/dd/yy)?

10. When did you make your last admissions offers before the standard acceptance deadline this year?

    ○ Mar 1-15
    ○ Mar 16-31
    ○ Apr 1-14
    ○ Apr 15
    ○ Other (please specify)

11. Did you make any offers after your standard acceptance deadline?

12. Do you routinely make offers after your standard acceptance deadline?

13. When do a significant number of admits notify you of their decision (check all that apply)?

    ☐ Before March 15
    ☐ From March 15 to March 31
    ☐ From April 1 to April 7
    ☐ From April 8 through April 13
    ☐ On April 14 and 15
    ☐ After April 15

AAS Graduate Admissions Department Survey

## Master's Applications

14. Do you accept Master's applications (to clarify, some programs grant a Masters en route to a PhD but do not admit students for a Masters-specific track. If this is the case for your program please answer "no")?

☐

## AAS Graduate Admissions Department Survey

### Masters Application Numbers and Timing

**Please note that we request information for the past 6 years in order to get information on trends that include one full year before the COVID-19 pandemic.**

15. How many Master's applications did you receive in the following academic years?

2023-2024: ☐
2022-2023: ☐
2021-2022: ☐
2020-2021: ☐
2019-2020: ☐
2018-2019: ☐

16. How many Master's offers did you make in each of these years?

2023-2024: ☐
2022-2023: ☐
2021-2022: ☐
2020-2021: ☐
2019-2020: ☐
2018-2019: ☐

17. How many Master's offers were accepted in each of these years?

2023-2024: ☐
2022-2023: ☐
2021-2022: ☐
2020-2021: ☐
2019-2020: ☐
2018-2019: ☐

18. When was your most recent admissions application deadline (mm/dd/yy)?

19. When did you make your first admissions offers this year?
- ○ Jan 1-15
- ○ Jan 16-31
- ○ Feb 1-15
- ○ Feb 16-29
- ○ Mar/Apr
- ○ Other (please specify)

20. When was your standard acceptance deadline (mm/dd/yy)?

21. When did you make your last admissions offers before the standard acceptance deadline this year?
- ○ Mar 1-15
- ○ Mar 16-31
- ○ Apr 1-14
- ○ Apr 15
- ○ Other (please specify)

22. Did you make any offers after your standard acceptance deadline?

23. Do you routinely make offers after your standard acceptance deadline?

24. When do a significant number of admits notify you of their decision (check all that apply)?
- ☐ Before March 15
- ☐ From March 15 to March 31
- ☐ From April 1 to April 7
- ☐ From April 8 through April 13
- ☐ On April 14 and 15
- ☐ After April 15

## AAS Graduate Admissions Department Survey

### Bridge Applications

25. Do you accept applications for a bridge program?

☐

## AAS Graduate Admissions Department Survey

### Bridge Application Numbers and Timing

**Please note that we request information for the past 6 years in order to get information on trends that include one full year before the COVID-19 pandemic.**

26. How many Bridge applications did you receive in the following academic years?

2023-2024: ______
2022-2023: ______
2021-2022: ______
2020-2021: ______
2019-2020: ______
2018-2019: ______

27. How many Bridge offers did you make in each of these years?

2023-2024: ______
2022-2023: ______
2021-2022: ______
2020-2021: ______
2019-2020: ______
2018-2019: ______

28. How many Bridge offers were accepted in each of these years?

2023-2024: ☐

2022-2023: ☐

2021-2022: ☐

2020-2021: ☐

2019-2020: ☐

2018-2019: ☐

29. When was your most recent admissions application deadline (mm/dd/yy)?

☐

30. When did you make your first admissions offers this year?

- ○ Jan 1-15
- ○ Jan 16-31
- ○ Feb 1-15
- ○ Feb 16-29
- ○ Mar/Apr
- ○ Other (please specify)

☐

31. When was your standard acceptance deadline (mm/dd/yy)?

☐

32. When did you make your last admissions offers before the standard acceptance deadline this year?

- ○ Mar 1-15
- ○ Mar 16-31
- ○ Apr 1-14
- ○ Apr 15
- ○ Other (please specify)

☐

33. Did you make any offers after your standard acceptance deadline?

☐

34. Do you routinely make offers after your standard acceptance deadline?

☐

35. When do a significant number of admits notify you of their decision (check all that apply)?

- [ ] Before March 15
- [ ] From March 15 to March 31
- [ ] From April 1 to April 7
- [ ] From April 8 through April 13
- [ ] On April 14 and 15
- [ ] After April 15

## AAS Graduate Admissions Department Survey

### Application Policies

**We are particularly interested in collecting information on application policies for PhD programs. However, some departments that accept applications for more than one type of program (PhD, Master's, Bridge) may have somewhat different application policies for different degree tracks. If this applies to your department and you accept applications for the PhD, please answer for your PhD program. If you do not have a PhD program but accept applications for separate Masters and/or Bridge programs and they have different policies, please specify which program you are answering for below.**

36. Are you answering the following questions regarding your PhD, Masters, or Bridge program?

37. How do department faculty [other than the dept. chair and graduate admissions chair(s)] and current graduate students communicate with potential applicants before they apply (check all that apply)?

- [ ] faculty actively communicate with potential applicants via email
- [ ] current graduate students actively communicate with potential applicants via email
- [ ] faculty actively communicate with potential applicants via zoom
- [ ] current graduate students actively communicate with potential applicants via zoom
- [ ] faculty communication with potential applicants is discouraged
- [ ] graduate student communication with potential applicants is discouraged
- [ ] potential applicants sometimes make in-person visits to the department before the application deadline
- [ ] in-person visits before the application deadline are discouraged

38. Do you have a formal policy on how faculty communicate with potential applicants?

39. Do you have a formal policy on how current graduate students communicate with potential applicants?

## AAS Graduate Admissions Department Survey

### Application Fees

40. Do you charge an application fee?

## AAS Graduate Admissions Department Survey

### Application Fee Details

41. What is the amount of your application fee?

42. Do you offer fee waivers?

## AAS Graduate Admissions Department Survey

### Application Components and Reviewing

Please answer the following questions thinking about how your department uses the requested information as opposed to whether or not the university requires it. For example, we are aware of some programs where the university requires submission of regular GRE scores but the department does not consider them. In that case, please answer the question about the regular GRE as "not used."

43. How does your program use the Physics GRE (check one)?

- ○ Required
- ○ Recommended
- ○ Optional
- ○ Not used

44. If you answered "not used" to the previous question, what year when did you stop using the physics GRE?

45. How does your program use the regular GRE (check one)?

- ○ Required
- ○ Recommended
- ○ Optional
- ○ Not used

46. If you answered "not used" to the previous question, what year did you stop using the regular GRE?

[          ]

47. How many letters of recommendation do you require?

[          ]

48. What statements do you require or request as optional and what are the lengths of these if specified (e.g. 1-page personal statement required, 1-page research statement required, 1-page diversity statement optional)?

[          ]

49. How many people total review applications in a typical year?

[          ]

50. What is the rank of the reviewers?

- ○ All tenured or tenure-track (T/TT) faculty
- ○ A mix of T/TT and non-tenure track faculty
- ○ A mix of faculty and non-faculty ranks
- ○ All non-faculty ranks

51. How many reviewers are expected to read each application?

[          ]

52. With regard to how applications are graded:

- ○ We use a checklist of qualifications
- ○ We use a specified grading rubric with points assigned to categories
- ○ We use an ungraded rubric/set of guidelines
- ○ We do not use defined rubric, guidelines, or checklist

AAS Graduate Admissions Department Survey

## Interviews

53. Do you conduct interviews of applicants (y/n)?

    [ ]

## AAS Graduate Admissions Department Survey

### Interview Details

54. When did you start interviewing as part of your application process (leave blank if you do not interview)?

    ( ) In the last 1-3 years
    ( ) 4-5 years ago
    ( ) More than 5 years ago

55. How are the interviews conducted (check all that apply)?

    [ ] Email
    [ ] Phone
    [ ] Zoom
    [ ] In person
    [ ] Other (please specify)

56. When do you typically interview (check all that apply)?

    [ ] Early January
    [ ] Late January
    [ ] Early February
    [ ] Late February
    [ ] Early March
    [ ] Late March
    [ ] Other (please specify)

## AAS Graduate Admissions Department Survey

### Application Grading Criteria

57. How important are the following factors in identifying top candidates in the admissions process?

| | Not Important | Slightly Important | Important | Very Important | N/A |
|---|---|---|---|---|---|
| Physics GRE Score | ○ | ○ | ○ | ○ | ○ |
| General GRE Score | ○ | ○ | ○ | ○ | ○ |
| Grades at undergraduate institution(s) | ○ | ○ | ○ | ○ | ○ |
| Reputation of undergraduate degree granting institution | ○ | ○ | ○ | ○ | ○ |
| Letters of recommendation | ○ | ○ | ○ | ○ | ○ |
| Reviewers' familiarity with at least some of the letter writers | ○ | ○ | ○ | ○ | ○ |
| Personal Statement/written materials | ○ | ○ | ○ | ○ | ○ |
| Student reaches out to faculty before/during application process | ○ | ○ | ○ | ○ | ○ |
| How student performed in an interview | ○ | ○ | ○ | ○ | ○ |
| Participation in an REU-style intensive (full-time) summer research program | ○ | ○ | ○ | ○ | ○ |
| First authorship on a submitted refereed paper | ○ | ○ | ○ | ○ | ○ |
| Co-authorship on a submitted refereed paper | ○ | ○ | ○ | ○ | ○ |

58. There is a perception among many applicants that authorship on a published or submitted refereed paper is required for admission to graduate school. For applicants to you department, please indicate the significance of authorship on a refereed paper (check one):

○ Almost all admitted students have at least one refereed publication

○ A majority of admitted students have at least one refereed publication

○ While some admitted students have a refereed publication, most do not

○ Very few admitted students have a refereed publication

## AAS Graduate Admissions Department Survey

### Visits for admitted students.

59. Do you offer admitted students the opportunity to visit your institution (y/n)?

[          ]

## AAS Graduate Admissions Department Survey

### Visit Details

60. When did you start offering visits?

- ○ In the last 1-3 years
- ○ 4-5 years ago
- ○ More than 5 years ago

61. If you do organized group visits, when do you typically hold these visits for admitted students?

- ○ Early January
- ○ Late January
- ○ Early February
- ○ Late February
- ○ Early March
- ○ Late March
- ○ Other (please specify)

[                                    ]

62. For admitted student visits, do you (check all that apply)

- ☐ Try to bring all visiting students in in one or two organized group visits
- ☐ Bring students in whenever works best for individual students

## AAS Graduate Admissions Department Survey

### Shared Application Platform

**Some fields already use a shared application platform for post graduate studies, kind of like the common app used by many schools for undergraduate admissions. In some case, there is an opportunity to add some limited customization to such shared platforms, such as asking for statement that explains why a candidate would like to attend your specific university.**

63. If you were allowed to make limited customizations, such as adding a request for a specific statement on why a candidate was interested in your program, would you be interested in using a shared application platform?

## AAS Graduate Admissions Department Survey

Open Answer/Feedback Questions

64. Over the past five years, what challenges have stood out in your graduate admissions process (e.g. number of applicants, class size, recent changes in affirmative action policies)?

65. Has your department changed aspects of the application process/review or taken steps to address these challenges, and if so, how?

66. Are there larger-scale solutions (beyond your department) that you would like to see implemented to address these challenges?